\documentclass[journal]{IEEEtran}
\ifCLASSINFOpdf
\else
\fi

\usepackage{stfloats}

\usepackage{cite}
\usepackage{amsmath,amssymb,amsfonts}
\usepackage{subcaption}
\usepackage{graphicx}
\usepackage{textcomp}
\usepackage{xcolor}

\usepackage{tikz, pgfplots}
\usepackage{filecontents}
\usepackage{pgfplotstable}
\usepgfplotslibrary{patchplots}
\usetikzlibrary{positioning}
\usetikzlibrary{calc,through,backgrounds}
\usetikzlibrary{shapes.misc}
\usetikzlibrary{shapes.geometric}
\usetikzlibrary{spy}

  \pgfplotsset{compat=newest}
  \usetikzlibrary{plotmarks}
  \usetikzlibrary{arrows.meta}
  \usepgfplotslibrary{patchplots}
  \usepackage{grffile}

\usepackage{enumerate}
\usepackage{multirow}
\usepackage[nohyperlinks, nolist]{acronym}
\usepackage{epstopdf}
\usetikzlibrary{arrows.meta}
\usepackage{float}
\usetikzlibrary{matrix}
\graphicspath{{./fig/}}
\def\BibTeX{{\rm B\kern-.05em{\sc i\kern-.025em b}\kern-.08em
    T\kern-.1667em\lower.7ex\hbox{E}\kern-.125emX}}

  \tikzset{cross/.style={cross out, draw=black, minimum size=2*(#1-\pgflinewidth), inner sep=0pt, outer sep=0pt},
cross/.default={2pt}}    
      
\usepackage[linesnumbered, ruled]{algorithm2e}
\SetKwRepeat{Do}{do}{while}%

\newcommand{\ve}{\mathbf}
\newcommand{\m}{\mathbf}

\newcommand{\vef}[1]{\mathbf{\widetilde{\mathbf{#1}}}} 

\newcommand{\veh}[1]{\widehat{\mathbf{#1}}}
\newcommand{\mh}[1]{\widehat{\mathbf{#1}}}

\newcommand*{\ReadOutElement}[4]{%
    \pgfplotstablegetelem{#2}{#3}\of{#1}%
    \let#4\pgfplotsretval
}

\newdimen\CdotAxis
\newcommand*{\CdotAux}[3]{%
  {%
    \settoheight\CdotAxis{$#2\vcenter{}$}%
    \sbox0{%
      \raisebox\CdotAxis{%
        \scalebox{#1}{%
          \raisebox{-\CdotAxis}{%
            $\mathsurround=0pt #2#3$%
          }%
        }%
      }%
    }%
    \dp0=0pt %
    \sbox2{$#2\bullet$}%
    \ifdim\ht2<\ht0 %
      \ht0=\ht2 %
    \fi
    \sbox2{$\mathsurround=0pt #2#3$}%
    \hbox to \wd2{\hss\usebox{0}\hss}%
  }%
}

\begin{document}
%


\begin{acronym}

\acro{aaf}[AAF]{anti-aliasing filter}
\acro{ad}[AD]{autonomous driving}
\acro{adc}[ADC]{analog-to-digital converter}
\acro{adas}[ADAS]{advanced driver assistance systems}
\acro{als}[ALS]{approximate least squares}
\acro{aoa}[AOA]{angle of arrival}
\acro{aod}[AOD]{angle of departure}
\acro{awgn}[AWGN]{additive white Gaussian noise}
\acro{bb}[BB]{baseband}
\acro{ber}[BER]{bit error ratio}
\acro{blue}[BLUE]{best linear unbiased estimator}
\acro{bmse}[BMSE]{Bayesian mean square error}
\acro{bpsk}[BPSK]{binary phase shift keying}
\acro{bwlue}[BWLUE]{best widely linear unbiased estimator}
\acro{ccdf}[CCDF]{complementary cumulative distribution function}
\acro{cfo}[CFO]{carrier frequency offset}
\acro{cfr}[CFR]{channel frequency response}
\acro{cir}[CIR]{channel impulse response}
\acro{cpe}[CPE]{common phase error}
\acro{cr}[CR]{corner reflector}
\acro{cs}[CS]{compressed sensing}
\acro{cwcu}[CWCU]{component-wise conditionally unbiased}
\acro{cl}[CWCU LMMSE]{component-wise conditionally unbiased linear minimum mean square error}
\acro{cp}[CP]{cyclic prefix}
\acro{cwl}[CWCU WLMMSE]{component-wise conditionally unbiased widely linear minimum mean square error}
\acro{dac}[DAC]{digital-to-analog converter}
\acro{dbf}[DBF]{digital beamforming}
\acro{dc}[DC]{direct current}
\acro{dft}[DFT]{discrete Fourier transform}
\acro{ddm}[DDM]{Doppler-division multiplexing}
\acro{dpsk}[DPSK]{differential phase shift keying}
\acro{ecir}[ECIR]{effective channel impulse response}
\acro{ecfr}[ECFR]{effective channel frequency response}
\acro{em}[EM]{expectation-maximization}
\acro{esi}[ESI]{equidistant subcarrier interleaving}
\acro{fmcw}[FMCW]{frequency-modulated continuous wave}
\acro{fft}[FFT]{fast Fourier transform}
\acro{fi}[FI]{frequency independent}
\acro{fir}[FIR]{finite impulse response}
\acro{fs}[FS]{frequency selective}
\acro{ici}[ICI]{inter-carrier interference}
\acro{isi}[ISI]{inter-symbol interference}
\acro{idft}[IDFT]{inverse discrete Fourier transform}
\acro{if}[IF]{intermediate frequency}
\acro{ifft}[IFFT]{inverse fast Fourier transform}
\acro{iid}[i.i.d.]{independent and identically distributed}
\acro{iq}[IQ]{in-phase and quadrature-phase}
\acro{iqir}[FRIQIR]{full-range IQ imbalance robust}
\acro{iqirSotA}[IQIR]{IQ imbalance robust}
\acro{llr}[LLR]{log-likelihood ratio}
\acro{lmmse}[LMMSE]{linear minimum mean square error}
\acro{lms}[LMS]{least mean square}
\acro{lna}[LNA]{low-noise amplifier}
\acro{los}[LOS]{line of sight}
\acro{lpf}[LPF]{low-pass filter}
\acro{ls}[LS]{least squares}
\acro{lti}[LTI]{linear time-invariant}
\acro{map}[MAP]{maximum a posteriori}
\acro{mimo}[MIMO]{multiple-input multiple-output}
\acro{miso}[MISO]{multiple-input single-output}
\acro{ml}[ML]{maximum likelihood}
\acro{mlem}[ML-EM]{maximum likelihood expectation-maximization}
\acro{mmse}[MMSE]{minimum mean square error}
\acro{mse}[MSE]{mean square error}
\acro{mvdr}[MVDR]{minimum variance distortionless response}
\acro{mvu}[MVU]{minimum variance unbiased}
\acro{nlms}[NLMS]{normalized least mean squares}
\acro{nlos}[NLOS]{non-line of sight}
\acro{ofdm}[OFDM]{orthogonal frequency-division multiplexing}
\acro{papr}[PAPR]{peak-to-average power ratio}
\acro{pdf}[PDF]{probability density function}
\acro{pll}[PLL]{phase-locked loop}
\acro{pwcu}[PWCU]{part-wise conditionally unbiased}
\acro{pwl}[PWCU WLMMSE]{part-wise conditionally unbiased widely linear minimum mean square error}
\acro{qam}[QAM]{quadrature amplitude modulation}
\acro{qpsk}[QPSK]{quadrature phase-shift keying}
\acro{rcs}[RCS]{radar cross section}
\acro{rdm}[RDMult]{range-division multiplexing}
\acro{rf}[RF]{radio frequency}
\acro{rls}[RLS]{recursive least squares}
\acro{rvm}[RDM]{range-Doppler map}
\acro{rx}[Rx]{receiver}
\acro{sa}[SA]{subcarrier aliasing}
\acro{sim}[SIM]{spectral interleaving multiplexing}
\acro{siso}[SISO]{single-input single-output}
\acro{snr}[SNR]{signal-to-noise ratio}
\acro{stln}[STLN]{structured total least norm}
\acro{stls}[STLS]{structured total least squares}
\acro{tls}[TLS]{total least squares}
\acro{tdm}[TDM]{time-division multiplexing}
\acro{tx}[Tx]{transmitter}
\acro{ula}[ULA]{uniform linear array}
\acro{uwofdm}[UW-OFDM]{unique-word orthogonal frequency division multiplexing}
\acro{wlan}[WLAN]{wireless local area network}
\acro{wlls}[WLLS]{widely linear least squares}
\acro{wlmmse}[WLMMSE]{widely linear minimum mean square error}
\acro{wls}[WLS]{weighted least squares}
\acro{wrt}[w.r.t.]{with respect to}
\acro{wwlls}[WWLLS]{weighted widely linear least squares}

\end{acronym}

\title{OFDM-based Waveforms for Joint Sensing and Communications Robust to Frequency Selective IQ Imbalance}
%
%
%

\author{Oliver~Lang,~\IEEEmembership{Member,~IEEE,}
		Christian~Hofbauer,~\IEEEmembership{Member,~IEEE,}
		Moritz~Tockner,~\IEEEmembership{Student~Member,~IEEE,}
        Reinhard~Feger,~\IEEEmembership{Member,~IEEE,}
        Thomas~Wagner,~\IEEEmembership{Member,~IEEE,}
        and~Mario~Huemer,~\IEEEmembership{Senior~Member,~IEEE}
\thanks{Oliver Lang is with the Institute
of Signal Processing, Johannes Kepler University, Linz,
Austria (e-mail: oliver.lang@jku.at).}
\thanks{Christian Hofbauer is with Silicon Austria Labs GmbH, Linz, Austria (e-mail: christian.hofbauer@silicon-austria.com).}
\thanks{Moritz Tockner is with the Christian Doppler Laboratory for Digitally Assisted RF Transceivers for Future Mobile Communications, Institute of Signal Processing, Johannes Kepler University, Linz, Austria (e-mail: moritz.tockner@jku.at).} 
\thanks{Reinhard Feger and Thomas Wagner are with the Institute for Communications Engineering and RF-Systems, Johannes Kepler University, Linz, Austria (e-mail: \{reinhard.feger; thomas.wagner\}@jku.at).}
\thanks{Mario Huemer is with the Institute
of Signal Processing, Johannes Kepler University, Linz,
Austria, and with the JKU LIT SAL eSPML Lab, 4040 Linz, Austria (e-mail: mario.huemer@jku.at).}
\thanks{ The work of Christian Hofbauer was supported by Silicon Austria Labs (SAL), owned by the Republic of Austria, the Styrian Business Promotion Agency (SFG), the federal state of Carinthia, the Upper Austrian Research (UAR), and the Austrian Association for the Electric and Electronics Industry (FEEI). 

This work was supported by the "University SAL Labs" initiative of SAL and its Austrian partner universities for applied fundamental research for electronic based systems. 

The financial support by the Austrian Federal Ministry for Digital and Economic Affairs, the National Foundation for Research, Technology and Development and the Christian Doppler Research Association are gratefully acknowledged.}
}

\maketitle

\begin{abstract}
\Ac{ofdm} is a promising waveform candidate for future joint sensing and communication systems. It is well known that the \ac{ofdm} waveform is vulnerable to \ac{iq} imbalance, which increases the noise floor in a \ac{rvm}. A state-of-the-art method for robustifying the \ac{ofdm} waveform against \ac{iq} imbalance avoids an increased noise floor, but it generates additional ghost objects in the \ac{rvm} \cite{IQ_Imbalance_Rad2}. A consequence of these additional ghost objects is a reduction of the maximum unambiguous range. In this work, a novel \ac{ofdm}-based waveform robust to \ac{iq} imbalance is proposed, which neither increases the noise floor nor reduces the maximum unambiguous range. The latter  is achieved by shifting the ghost objects in the \ac{rvm} to different velocities such that their range variations observed over several consecutive \acp{rvm} do not correspond to the observed velocity. This allows tracking algorithms to identify them as ghost objects and eliminate them for the follow-up processing steps. Moreover, we propose complete communication systems for both the proposed waveform as well as for the state-of-the-art waveform, including methods for channel estimation, synchronization, and data estimation that are specifically designed to deal with frequency selective \ac{iq} imbalance which occurs in wideband systems. The effectiveness of these communication systems is demonstrated by means of \ac{ber} simulations. 
\end{abstract}


\acresetall

%
\IEEEpeerreviewmaketitle

\section{Introduction}  \label{sec:Introduction}

\IEEEPARstart{J}{oint} sensing and communications is in the focus of intensive research due to the benefits of using the same hardware, waveform, and bandwidth for two tasks: sensing the environment, and communicating information. While various forms of joint sensing and communication architectures exist, the focus of this work lies on so-called dual-function radar-communication systems that use the very same transmit signals for both tasks simultaneously. Potential applications include, e.g., automotive vehicle-to-vehicle communication, and environment sensing for mobile devices and smartphones \cite{akan2020internet, Sit_Automotive_MIMO_OFDM, Braun_Parametrization}. A prominent waveform candidate for such systems is \ac{ofdm}. The \ac{ofdm} waveform is well established from communication applications \cite{Van_Nee_OFDM, salehi2007digital, haykin2013digital}, while its utilization for radar sensing is a topic of intensive research \cite{Levanon_Multifrequency_complementary, Donnet_Combining_MIMO_Radar, Sturm_A_novel_approach, Hakobyan_Inter_Carrier_Interference, Sit_Automotive_MIMO_OFDM, Sturm_An_OFDM_System,  Lang_Asilomar_2020, Lang_SA_OFDM, Lang_RDM_JP}.

The hardware for transmitting and receiving \ac{ofdm} signals is of course not perfect in practice and may introduce unwanted signal distortions, e.g., due to \ac{iq} imbalance. For communication applications,  \ac{iq} imbalance causes \ac{ici}, which degrades the \ac{ber} performance. Many \ac{iq} imbalance estimation and cancellation algorithms have been proposed \cite{IQ_Imbalance_Comm1, IQ_Imbalance_Comm2, IQ_Imbalance_Comm3,  IQ_Imbalance_Comm5}, whereas most of these methods exploit the circularity of a signal \cite{Second-order_analysis_of_improper_complex_random_vectors_and_processes, Statistical-Signal-Processing-of-Complex-Valued-Data-The-Theory-of-Improper-and-Noncircular-Signals, Essential_Statistics_and_Tools_for_Complex_Random_Variables, Diss_Lang_Oliver}. We refer to Sec.~\ref{sec:SOTA} for a more detailed review of those methods.

For radar applications, \ac{iq} imbalance translates to an increased noise level in the \ac{rvm} as well as the occurence of ghost objects \cite{IQ_Imbalance_Rad1}. Whether these ghost objects are visible in the \ac{rvm} or not depends on the \ac{snr}. 
There exist two common approaches in the literature to tackle these issues. The first approach is to estimate and cancel the \ac{iq} imbalance in the radar receiver as demonstrated in \cite{IQ_Imbalance_Rad1}. The method proposed therein relies on the assumption of circular receive signals. Unfortunately, this assumption does not hold in the presence of strong stationary objects, which motivated the investigation of a modified \ac{iq} imbalance estimation procedure  based on the cancellation of strong stationary objects in the \ac{rvm} \cite{IQ_Imbalance_Rad1}. 

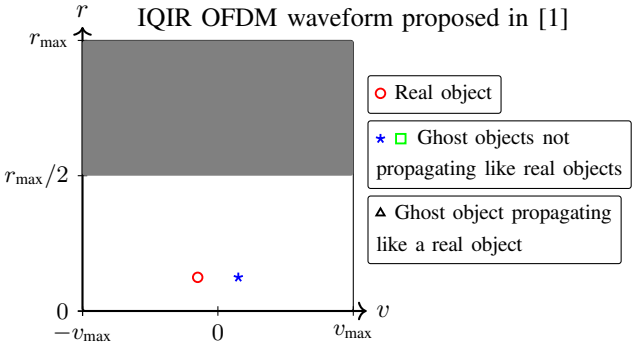
\begin{figure}[!t]
\centering
\begin{tikzpicture}[scale=0.9, rounded corners=1pt,inner sep=3.2pt,node distance=.2cm,every text node part/.style={align=left}]
	\node[above] at (4,4) {IQIR OFDM waveform proposed in \cite{IQ_Imbalance_Rad2}};

    \draw [<->,thick] (0,4.2) node (yaxis) [above] {$r$}
        |- (4.2,0) node (xaxis) [right] {$v$};
    \draw[draw = black] (0pt,2pt) -- (0pt,-2pt) node[below] {\small $-v_\text{max}$};
    \draw[draw = black] (2,2pt) -- (2,-2pt) node[below] {\small $0$};
    \draw[draw = black] (4,2pt) -- (4,-2pt) node[below] {\small $v_\text{max}$};
    \draw[draw = black] (2pt,0) -- (-2pt,0) node[left] {\small $0$};
    \draw[draw = black] (2pt,4) -- (-2pt,4) node[left] {\small $r_\text{max}$};
    \draw[draw = black] (2pt,2) -- (-2pt,2) node[left] {\small $r_\text{max}/2$};
    \draw[draw=black] (0,0) rectangle ++(4,4);
    %
    
\begin{axis}[
    anchor=origin,  
    x=1cm, y=1cm,   
    hide axis
]
\addplot [mark=+, color=black] table {
0 0
0 0
};

\addplot [only marks,thick, mark=o, color=red] coordinates {(1.7,0.5)};
\addplot [only marks,thick, mark=star, color=blue] coordinates {(2.3,0.5)};
\addplot [only marks,thick, mark=triangle, color=black] coordinates {(1.7,2.5)};
\addplot [only marks,thick, mark=square, color=green] coordinates {(2.3,2.5)};

\addplot [only marks,thick, mark=o, color=red] coordinates {(4.4,3.22)};
\addplot [only marks,thick, mark=star, color=blue] coordinates {(4.4,2.55)};
\addplot [only marks,thick, mark=triangle, color=black] coordinates {(4.4,1.45)};
\addplot [only marks,thick, mark=square, color=green] coordinates {(4.7,2.55)};
\end{axis}

\draw[draw = none, fill=gray, fill opacity=0.2, draw opacity=0] (0,2) rectangle ++(4,2);

\node[draw,minimum height = 0.5cm, minimum width = 0.5cm](state0) at (5.2,3.2) { $\hspace{1.2mm}$  \footnotesize Real object};

\node[draw, below=0.8cm of state0.west,minimum height = 0.5cm, minimum width = 0.5cm, anchor=west](state1) { $\hspace{4.2mm}$    \footnotesize Ghost objects not \\   \footnotesize propagating like real objects};

\node[draw, below=1.0cm of state1.west,minimum height = 0.5cm, minimum width = 0.5cm, anchor=west](state2) { $\hspace{1.2mm}$   \footnotesize Ghost object propagating\\  \footnotesize like a real object};

%

\end{tikzpicture}
\caption{RDM for the IQIR OFDM waveform in case of IQ imbalance in the transmitter and the receiver. This waveform avoids an increased noise floor due to IQ imbalance, but it generates 3 ghost objects for every real object in the RDM.}
\label{fig_SotA}
\end{figure}

The second approach, on which the proposed method in this work is built on, is to choose the subcarrier symbols in a way that makes the \ac{ofdm} waveform robust to \ac{iq} imbalance by design. This can, e.g., be achieved by applying a so-called $\pi / K$ phase modulation \cite{bauduin2022pi} onto the transmitted \ac{ofdm} symbols, followed by a coherent accumulation of multiple received \ac{ofdm} symbols in the radar receiver. Although this approach effectively reduces the degrading effects of \ac{iq} imbalance in the range spectrum, the coherent accumulation step requires repeatedly  transmitting the same \ac{ofdm} symbol, limiting the achievable data rate in  communications for joint sensing and communication applications. Hence, this waveform seems to be better suited for pure radar sensing applications.

Another design rule to robustify the \ac{ofdm} waveform is described in \cite{IQ_Imbalance_Rad2}, and the resulting waveform is referred to as \ac{iqirSotA} \ac{ofdm} in this work\footnote{A radar system or a joint sensing and communication system utilizing the \ac{iqirSotA} \ac{ofdm} waveform is simply referred to as \ac{iqirSotA} \ac{ofdm}. Similar short names are also used for other waveforms considered in this work.}. 
While a detailed discussion on this waveform in combination with \ac{tx} and \ac{rx} \ac{iq} imbalance is provided later in Sec.~\ref{sec:SOTA}, we note that the basic underlying concept to obtain robustness against \ac{iq} imbalance relies on introducing redundancy along the subcarrier symbols. This is obtained by dividing the set of subcarriers into two halves, with the first one available for arbitrary loading with, e.g., data payload, and the other half with redundancy generated systematically from the first one.
A radar system utilizing an \ac{iqirSotA} \ac{ofdm} waveform does not suffer from an \ac{iq} imbalance induced increase of the noise level, but instead, up to three ghost objects appear in the \ac{rvm} for every real object\footnote{In this work, we assume \ac{iq} imbalance in the transmitter and the receiver of a radar system, which leads to the mentioned 3 ghost objects per real object. In case \ac{iq} imbalance appears only in the receiver as assumed in \cite{IQ_Imbalance_Rad2}, only a single ghost object per real object would appear.}. This scenario is sketched in the \ac{rvm} in Fig.~\ref{fig_SotA}, where the red circle represents the real object and the other marks ghost objects. Two of these ghost objects (blue star and green square) appear at a different velocity bin than the corresponding real object. Consequently, the change in range observed along several consecutive \acp{rvm} do not correspond to the observed velocity. Hence, it will be assumed in this work that tracking algorithms \cite{bar1995multitarget, bar2011tracking, blackman1999design, 1263228} are able to identify and cancel these two ghost objects within a short period of time. The third ghost object (black triangle), is located at a different range bin but at the same velocity bin as the corresponding real object. Thus, its range variation observed over several consecutive \acp{rvm} cannot be used to identify it as a ghost object. 
This issue can be resolved by reducing the maximum unambiguous range $r_\text{max}$ by a factor of 2 (gray shaded area) \cite{IQ_Imbalance_Rad2}. Depending on the application and the parametrization of the \ac{ofdm} waveform, this reduction of $r_\text{max}$ may be a significant drawback, especially in combination with multiplexing techniques that reduce $r_\text{max}$ even further such as \ac{esi} \cite{Sit_Automotive_MIMO_OFDM, sturm2013spectrally} and \ac{rdm} \cite{Lang_RDM_JP}. It is worth noting that the \ac{iqirSotA} \ac{ofdm} waveform is suitable for joint sensing and communication applications, however, the communication aspects of this waveform were not analyzed in \cite{IQ_Imbalance_Rad2}.

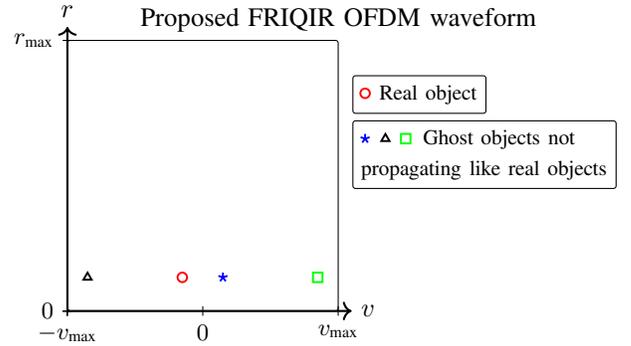
\begin{figure}[!t]
\centering
\begin{tikzpicture}[scale=0.9, rounded corners=1pt,inner sep=3.2pt,node distance=.2cm,every text node part/.style={align=left}]
	\node[above] at (4,4) {Proposed FRIQIR OFDM waveform};

    \draw [<->,thick] (0,4.2) node (yaxis) [above] {$r$}
        |- (4.2,0) node (xaxis) [right] {$v$};
    \draw[draw = black] (0pt,2pt) -- (0pt,-2pt) node[below] {\small $-v_\text{max}$};
    \draw[draw = black] (2,2pt) -- (2,-2pt) node[below] {\small $0$};
    \draw[draw = black] (4,2pt) -- (4,-2pt) node[below] {\small $v_\text{max}$};
    \draw[draw = black] (2pt,0) -- (-2pt,0) node[left] {\small $0$};
    \draw[draw = black] (2pt,4) -- (-2pt,4) node[left] {\small $r_\text{max}$};
    \draw[draw=black] (0,0) rectangle ++(4,4);
    
\begin{axis}[
    anchor=origin,  
    x=1cm, y=1cm,   
    hide axis
]
\addplot [mark=+, color=black] table {
0 0
0 0
};

\addplot [only marks,thick, mark=o, color=red] coordinates {(1.7,0.5)};
\addplot [only marks,thick, mark=star, color=blue] coordinates {(2.3,0.5)};
\addplot [only marks,thick, mark=triangle, color=black] coordinates {(0.3,0.5)};
\addplot [only marks,thick, mark=square, color=green] coordinates {(3.7,0.5)};

\addplot [only marks,thick, mark=o, color=red] coordinates {(4.4,3.22)};
\addplot [only marks,thick, mark=star, color=blue] coordinates {(4.4,2.55)};
\addplot [only marks,thick, mark=triangle, color=black] coordinates {(4.7,2.55)};
\addplot [only marks,thick, mark=square, color=green] coordinates {(5.0,2.55)};
\end{axis}


    %
%

\node[draw,minimum height = 0.5cm, minimum width = 0.5cm](state0) at (5.2,3.2) { $\hspace{1.2mm}$  \footnotesize Real object};

\node[draw, below=0.8cm of state0.west,minimum height = 0.5cm, minimum width = 0.5cm, anchor=west](state1) { $\hspace{7.0mm}$   \footnotesize Ghost objects not\\  \footnotesize propagating like real objects};

%

\end{tikzpicture}
\caption{RDM for the proposed FRIQIR OFDM waveform in case of IQ imbalance in the transmitter and the receiver.}
\label{fig_IQIR_OFDM}
\end{figure}

In this work, a novel \ac{ofdm}-based waveform for joint sensing and communications, by means of waveform generation at the transmitter and processing chains at the radar receiver and the communication receiver, is presented. This waveform is referred to as \ac{iqir} \ac{ofdm} and, as for \ac{iqirSotA} \ac{ofdm}, a robustness against \ac{iq} imbalance is achieved by adding redundancy to the transmit data. 
However, this is done in a different way, yielding the following benefits over \ac{iqirSotA} \ac{ofdm} \acused{wrt} \ac{wrt} radar sensing:
\begin{itemize}
\item All additional ghost objects are located in the \ac{rvm} at the same range bin as the real object, but at different velocity bins (see Fig.~\ref{fig_IQIR_OFDM}). Hence, their range variations observed along several consecutive \acp{rvm} diverge from those of a real object, since it does not correspond to the observed velocity. This enables tracking algorithms to identify them as ghost objects.
\item No reduction of the maximum unambiguous range $r_\text{max}$ is necessary.
\end{itemize}
Moreover, we compare \ac{iqir} against \ac{iqirSotA} also \ac{wrt} communications, focusing on the following contributions:
\begin{itemize}
\item The so-called \textit{effective channels}, a combination of the wireless propagation channel and the distortions caused by \ac{iq} imbalance, are analyzed for both waveforms and compared to the effective channel for the standard \ac{ofdm} waveform. For the proposed \ac{iqir} \ac{ofdm} waveform it turns out that even for static channels, we observe two different alternating effective channels due to \ac{iq} imbalance.    
\item Suitable methods for channel estimation, synchronization, and data estimation are proposed for the \ac{iqir} \ac{ofdm} waveform. However, since the communication aspects of \ac{iqirSotA} \ac{ofdm} were not analyzed in \cite{IQ_Imbalance_Rad2}, we propose such methods for this waveform as well, enabling a direct comparison between them. These methods are specifically designed to deal with the observed effective channels and to utilize the redundancy in the transmitted data efficiently. 
\item The effectiveness of the proposed methods for \ac{iqirSotA} \ac{ofdm} and \ac{iqir} \ac{ofdm} are demonstrated in form of \ac{ber} simulations, revealing a significant performance boost compared to standard \ac{ofdm}.
\end{itemize}

This paper is organized as follows: The employed model for \ac{iq} imbalance is described in Sec.~\ref{sec:Effects_IQ_Imbalance}. Sec.~\ref{sec:SOTA} presents the state of the art on \ac{iq} imbalance mitigation techniques for \ac{ofdm} radar applications. Sec.~\ref{sec:IQIR_OFDM_Radar} contains the derivation of the \ac{iqir} \ac{ofdm} waveform. Sec.~\ref{sec:Testing_SIM} presents simulation results in the context of radar sensing. The communication aspects of the considered waveforms are discussed in Sec.~\ref{sec:Comm_System}, and the 
\ac{ber} simulation results are shown in  Sec.~\ref{sec:BER_performance}. Finally, Sec.~\ref{sec:Conclusions} concludes this work.
\\
\emph{Notation}: 
\\ 
Lower-case bold face variables ($\ve{a}$, $\ve{b}$,...) indicate vectors, and upper-case bold face variables ($\m{A}$, $\m{B}$,...) indicate matrices. We further use $\mathbb{R}$ and $\mathbb{C}$ to denote the set of real- and complex-valued numbers, respectively, $(\cdot)^T$ to denote transposition, $(\cdot)^H$ to denote conjugate transposition, and $(\cdot)^*$ to denote complex conjugation. $\m{I}^{n}$ denotes the identity matrix of size $n\times n$, and $\m{0}^{n \times m}$ denotes the zero matrix of size $n\times m$. $\ve{1}^n$ indicates a column vector of length $n$ with all elements being one. $\oslash$ represents the Hadamard  division, and $\mathrm{j}$ represents the imaginary unit.  

\section{IQ Imbalance Model and its Effects on the RDM}  \label{sec:Effects_IQ_Imbalance}

This section discusses the employed \ac{iq} imbalance model and the effects of \ac{iq} imbalance on the  \acp{rvm}.


\subsection{IQ Imbalance Model for OFDM Signals}  \label{sec:Effects_IQ_Imbalance_A}

The basic \ac{ofdm} signal model for radar application as, e.g., described in \cite{Hakobyan_Inter_Carrier_Interference}, is assumed to be known by the reader. The number of subcarriers is denoted as $N_\text{c}$, and the number of consecutively transmitted \ac{ofdm} symbols is given by $N_\text{sym}$.  $N_\text{c}$ is assumed to be an even number. Furthermore, $B$ denotes the bandwidth, $f_\text{c}$ is the carrier frequency, $T = 1 / \Delta f$ is the duration of one \ac{ofdm} symbol, $\Delta f = B / N_\text{c}$ represents the subcarrier spacing, and $T_\text{cp}$ and $N_\text{cp}=T_\text{cp}\, B$ is the duration and the number of samples of the \ac{cp}, respectively.

Let $s_{k,m} \in \mathbb{C}$ denote the  subcarrier symbol, e.g., drawn from a \ac{qpsk} symbol alphabet, for the $k$th subcarrier of the $m$th \ac{ofdm} symbol. The subcarriers are indexed with $-N_\text{c}/2 \leq k < N_\text{c}/2$ with $k=0$ indicating \ac{dc}, and the \ac{ofdm} symbols are indexed with $0 \leq m < N_\text{sym}$. The subcarrier symbols distorted by \ac{iq} imbalance in the transmitter are denoted as $x_{k,m} \in \mathbb{C}$ yielding the relationship \cite{brotje2002estimation} 
 \begin{equation}
  x_{k,m} =\alpha^\text{Tx}_{k} \, s_{k,m} + \beta^\text{Tx}_{k} \, s^*_{-k,m}. \label{equ:IQIR_OFDM_005}      
\end{equation}
Please note that since $N_\text{c}$ is an even number, \eqref{equ:IQIR_OFDM_005} only holds for all values of $k$ when setting $s^*_{N_\text{c}/2,m} = 0$. Please note that practical implementations usually require a guard band implemented by setting the subcarrier symbols for several subcarriers at the band edges to zero, eliminating the need to model the \ac{iq} imbalance at $k=-N_\text{c}/2$ with \eqref{equ:IQIR_OFDM_005}. However, to make the proposed \ac{iqir} \ac{ofdm} comparable with \ac{iqirSotA} \ac{ofdm} in \cite{IQ_Imbalance_Rad2}, we do not consider a guard band in this work. 



Note that $\alpha^\text{Tx}_{k} \in \mathbb{C} $ and $\beta^\text{Tx}_{k} \in \mathbb{C}$ represent \ac{fs} \ac{iq} imbalance, since they depend on $k$, and may be derived as \cite{IQ_Imbalance_Comm2}
\begin{align}	
\alpha^\text{Tx}_{k} ={}& \text{cos}\left( \phi^\text{Tx}_{k} \right) + \text{j}\, \epsilon^\text{Tx}_{k} \, \text{sin}\left( \phi^\text{Tx}_{k}  \right)  \label{equ:IQIR_OFDM_004b}  \\
\beta^\text{Tx}_{k} ={}& \epsilon^\text{Tx}_{k}\, \text{cos}\left( \phi^\text{Tx}_{k} \right) + \text{j} \, \text{sin}\left( \phi^\text{Tx}_{k} \right) , \label{equ:IQIR_OFDM_004c}
\end{align}
with $\epsilon^\text{Tx}_{k} \in \mathbb{R}$ and $\phi^\text{Tx}_{k} \in \mathbb{R}$ denoting the amplitude and the phase mismatch for the $k$th subcarrier, respectively. Note that perfect \ac{iq} balance corresponds to $\alpha^\text{Tx}_{k} =1 $ and $\beta^\text{Tx}_{k}=0$.

The distorted subcarrier symbols $x_{k,m}$ in \eqref{equ:IQIR_OFDM_005} are transmitted, reflected by potentially multiple objects, and received again. With the frequency domain channel coefficients $c_{k,m} \in \mathbb{C}$, the received subcarrier symbols are given as
 \begin{align}
r_{k,m} ={}& c_{k,m}\, x_{k,m}  \in \mathbb{C}\\ 
={}& c_{k,m}\,\alpha^\text{Tx}_{k} \, s_{k,m} + c_{k,m}\,\beta^\text{Tx}_{k} \, s^*_{-k,m}  \label{equ:IQIR_OFDM_006}      
\end{align}
Additive noise and \ac{ici} \cite{Hakobyan_Inter_Carrier_Interference} are neglected in \eqref{equ:IQIR_OFDM_006} for the sake of readability, since they do not affect the derivation of the proposed waveform.  
Next, the received subcarrier symbols in \eqref{equ:IQIR_OFDM_006} are distorted by \ac{iq} imbalance in the receive path, modeled as
\begin{align}
y_{k,m} ={}& \alpha^\text{Rx}_{k} \, r_{k,m} + \beta^\text{Rx}_{k} \, r^*_{-k,m}  \in \mathbb{C} \\
={}& \alpha^\text{Rx}_{k} \, c_{k,m}\, \alpha^\text{Tx}_{k} \, s_{k,m}  + \alpha^\text{Rx}_{k} \, c_{k,m}\, \beta^\text{Tx}_{k} \, s^*_{-k,m} \nonumber \\
& \hspace{-3mm} + \beta^\text{Rx}_{k} \, c^*_{-k,m}\, \left(\alpha^\text{Tx}_{-k}\right)^* \, s^*_{-k,m} + \beta^\text{Rx}_{k} \, c^*_{-k,m}\, \left(\beta^\text{Tx}_{-k}\right)^* \, s_{k,m}
 \label{equ:IQIR_OFDM_007} 
\end{align}
for $-N_\text{c}/2 \leq k < N_\text{c}/2$, while the ideal case with perfect \ac{iq} balance is given by
\begin{equation}
 y_{k,m}^\text{ideal} = c_{k,m}\, s_{k,m}  \in \mathbb{C} \label{equ:OFDM_004qwert}
\end{equation}



\subsection{Effects of IQ Imbalance on the RDM}  \label{sec:Effects_IQ_Imbalance_B}

A matrix containing all subcarrier symbols $s_{k,m}$ is given in form of $\m{S} \in \mathbb{C}^{N_\text{c} \times N_\text{sym}}$. Each row of $\m{S}$ represents a subcarrier, and each column of $\m{S}$ represents an \ac{ofdm} symbol in frequency domain. 
Similarly, a matrix containing all received subcarrier symbols $y_{k,m}$ in \eqref{equ:IQIR_OFDM_007}  is given by $\m{Y}_\text{f,ts} \in \mathbb{C}^{N_\text{c} \times N_\text{sym}}$, where 'f' and 'ts' indicate frequency domain and  slow-time as the first and second dimensions. 
Based on $\m{Y}_\text{f,ts}$, the standard \ac{ofdm} receiver processing chain \cite{Sturm_A_novel_approach, Sturm_An_OFDM_System} is as follows:
\begin{itemize}
\item Performing an element-wise division $\m{Z}_\text{f,ts} = \m{Y}_\text{f,ts} \oslash \m{S}$.
\item Applying \ac{idft} operations on the columns of $\m{Z}_\text{f,ts}$ to obtain the range information, yielding $\m{Z}_\text{r,ts}$.
\item Applying \ac{dft} operations on the rows of $\m{Z}_\text{r,ts}$ to obtain the velocity information, yielding the desired \ac{rvm} $\m{Z}_\text{r,v}$. 
\end{itemize}
The maximum unambiguous range and the maximum unambiguous velocity of the \ac{rvm} are denoted as $r_\text{max}$ and $v_\text{max}$, respectively, and are given by 
\begin{align}
 r_\text{max} &= \frac{ c_0 N_\text{c} }{2 B},     & v_\text{max} &= \pm \frac{ c_0 }{4 f_\text{c}  \left(T + T_\text{cp} \right)},  \label{equ:OFDM_004}
\end{align}
 where $c_0$ is the speed of light.


For reasons of compactness, the effects of \ac{iq} imbalance on the \ac{rvm} are discussed based on \eqref{equ:IQIR_OFDM_007} without a complete derivation of the \ac{rvm}. 

A comparison with the ideal signal  $y_{k,m}^\text{ideal}$ in \eqref{equ:OFDM_004qwert} with perfect \ac{iq} balance reveals that the first term in \eqref{equ:IQIR_OFDM_007} is the ideal signal multiplied by the factor $\alpha^\text{Rx}_{k} \, \alpha^\text{Tx}_{k}$. This factor may cause distortions along the range axis in the \ac{rvm} especially when it varies heavily for different $k$. The second and third terms in \eqref{equ:IQIR_OFDM_007} represent interferences from the image subcarrier with subcarrier symbol $s^*_{-k,m}$. Since the subcarrier symbols are usually random, these interferences are also random and translate to an increase of the noise floor in the \ac{rvm}. The fourth term in \eqref{equ:IQIR_OFDM_007} contains the wanted subcarrier symbol $s_{k,m}$ and the complex-conjugate channel coefficient $c^*_{-k,m}$ from the image subcarrier. Due to the complex conjugation, this term may cause a ghost object for every real object at the opposite velocity in the \ac{rvm}.

\section{State of the Art}  \label{sec:SOTA}

As mentioned in Sec.~\ref{sec:Introduction}, two common approaches for tackling \ac{iq} imbalance exist, which are discussed in the following.

\subsection{Estimation/Cancellation of IQ Imbalance} 
In communications, \ac{iq} imbalance is a well-studied phenomenon and several methods for estimation and cancellation have been proposed. 
These include pilot- and preamble-based  \cite{IQ_Imbalance_Comm6, IQ_Imbalance_Comm8,IQ_Imbalance_Comm9, IQ_Imbalance_Comm2, IQ_Imbalance_Comm10, IQ_Imbalance_Comm11}, as well as statistics-based methods \cite{IQ_Imbalance_Comm3, paireder2019ultra, paireder2021enhanced, anttila2012blind, Diss_Michael_Petit, petit2015analysis}.
However, typical assumptions such as perfect channel knowledge \cite{IQ_Imbalance_Comm1}, \ac{fi} \ac{iq} imbalance \cite{IQ_Imbalance_Comm1, IQ_Imbalance_Comm10, haq2015correction}, circular signals \cite{IQ_Imbalance_Comm2}, \ac{iq} imbalance in the transmitter only \cite{IQ_Imbalance_Comm5, IQ_Imbalance_Comm7, brotje2002estimation}, a special transmitter design \cite{Fettweis_IQ}, or quasi-static channels \cite{IQ_Imbalance_Comm6} cause many existing methods to be unsuited for radar applications \cite{IQ_Imbalance_Rad1}. 



 The statistics-based methods usually assume the ideal time domain receive signal to be circular \cite{ Statistical-Signal-Processing-of-Complex-Valued-Data-The-Theory-of-Improper-and-Noncircular-Signals, Diss_Lang_Oliver}, and that the time domain receive signal becomes non-circular due to \ac{iq} imbalance (with \cite{paireder2019ultra} as an exception, which instead requires a uniform phase distribution). 
In radar applications, strong stationary objects or direct coupling cause the time domain receive signal to be non-circular as well \cite{IQ_Imbalance_Rad1}. Hence, many statistics-based methods for \ac{iq} imbalance estimation may fail for radar applications. 

A statistics-based method applicable to radar applications is proposed in \cite{IQ_Imbalance_Rad1}. The first step of this method is to derive the \ac{rvm} without any \ac{iq} imbalance cancellation. The second step is to remove strong stationary objects in the \ac{rvm}, which causes the receive signal to be non-circular. The result is then transformed back to time domain and fed into the statistics-based method from \cite{IQ_Imbalance_Comm3} to estimate and cancel \ac{iq} imbalance. 
The reader is referred to \cite{IQ_Imbalance_Rad1} for more details.




\subsection{Robust Waveform Design} 
For this approach, which is also the underlying principle for the waveform proposed in this work, the subcarrier symbols $s_{k,m}$ are chosen in a way that makes the \ac{ofdm} waveform robust to \ac{iq} imbalance by means of avoiding an \ac{iq} imbalance-induced increase of the noise floor. An interesting waveform referred to as \ac{iqirSotA} \ac{ofdm} is described in \cite{IQ_Imbalance_Rad2, IQ_Imbalance_Rad3} and in a related form also in \cite{lopez2006optimal}. The design procedure for the subcarrier symbols $s_{k,m}$ described therein to achieve the desired robustness against \ac{iq} imbalance works as follows.

\noindent \textit{DESIGN RULE I:}
\begin{enumerate}
\item The subcarrier symbols $s_{k,m}$ for $ 0 < k < N_\text{c}/2$ are chosen arbitrarily (e.g., by loading with data to be transmitted).
\item The subcarrier symbols $s_{k,m}$ for $k \in \{ -N_\text{c}/2,\,\, 0  \}$ are chosen to be real-valued.
\item The magnitude values of the subcarrier symbols $s_{k,m}$ for $ -N_\text{c}/2 < k < 0$ satisfy
\begin{equation}
|s_{k,m}| = |s_{-k,m}|. \label{equ:IQIR_OFDM_027d} 
\end{equation}
\item The phases of the subcarrier symbols $s_{k,m}$ for $ -N_\text{c}/2 < k < 0$ fulfill
\begin{equation}
\angle s_{k,m} + \angle s_{-k,m} = k\, \pi. \label{equ:IQIR_OFDM_027e} 
\end{equation}
\end{enumerate}
Combining \eqref{equ:IQIR_OFDM_027d} and \eqref{equ:IQIR_OFDM_027e} yields
\begin{equation}
s_{k,m} = s_{-k,m}^*  \, \text{e}^{\text{j} \pi k}, \label{equ:IQIR_OFDM_027a} 
\end{equation}
which is valid for $ -N_\text{c}/2  < k < N_\text{c}/2$.
The radar receiver processing chain for this waveform is the same as for standard \ac{ofdm} discussed in Sec.~\ref{sec:Effects_IQ_Imbalance}. A radar system based on \ac{iqirSotA} \ac{ofdm} was tested in \cite{IQ_Imbalance_Rad2} with \ac{iq} imbalance in the receiver but with an ideal transmitter. In this work, both sources of \ac{iq} imbalance as described in \eqref{equ:IQIR_OFDM_007} are considered. 
Inserting \eqref{equ:IQIR_OFDM_027a} into \eqref{equ:IQIR_OFDM_007} leads to
\begin{align}
y_{k,m} ={}& \alpha^\text{Rx}_{k} \, c_{k,m}\, \alpha^\text{Tx}_{k} \, s_{k,m} \nonumber \\
& + \alpha^\text{Rx}_{k} \, c_{k,m}\, \beta^\text{Tx}_{k} \, s_{k,m} \, \text{e}^{-\text{j} \pi k} \nonumber \\
& + \beta^\text{Rx}_{k} \, c^*_{-k,m}\, \left(\alpha^\text{Tx}_{-k}\right)^* \, s_{k,m} \, \text{e}^{-\text{j} \pi k} \nonumber \\
& + \beta^\text{Rx}_{k} \, c^*_{-k,m}\, \left(\beta^\text{Tx}_{-k}\right)^* \, s_{k,m}.
 \label{equ:IQIR_OFDM_028} 
\end{align}
Dividing $y_{k,m}$ in \eqref{equ:IQIR_OFDM_028} by $s_{k,m}$ results in
\begin{align}
z_{k,m} ={}& \alpha^\text{Rx}_{k} \, c_{k,m}\, \alpha^\text{Tx}_{k} \nonumber \\
& + \alpha^\text{Rx}_{k} \, c_{k,m}\, \beta^\text{Tx}_{k}  \, \text{e}^{-\text{j} \pi k} \nonumber \\
& + \beta^\text{Rx}_{k} \, c^*_{-k,m}\, \left(\alpha^\text{Tx}_{-k}\right)^* \, \text{e}^{-\text{j} \pi k} \nonumber \\
& + \beta^\text{Rx}_{k} \, c^*_{-k,m}\, \left(\beta^\text{Tx}_{-k}\right)^*,
 \label{equ:IQIR_OFDM_029} 
\end{align}
which can now be connected with the \ac{rvm} in Fig.~\ref{fig_SotA}. The first term in \eqref{equ:IQIR_OFDM_029} contains the wanted channel coefficient $c_{k,m}$ and it represents the real object (red circle) in Fig.~\ref{fig_SotA}. The additional scaling terms $\alpha^\text{Rx}_{k} \, \alpha^\text{Tx}_{k}$ may cause an increased sidelobe level along the range axis. The second term in \eqref{equ:IQIR_OFDM_029} includes the wanted channel coefficient $c_{k,m}$ modulated by $\text{e}^{-\text{j} \pi k}$. This modulation shifts this term by a distance of $r_\text{max}/2$ along the range axis, where it appears as a ghost object at the location of the black triangle. The third term in \eqref{equ:IQIR_OFDM_029} contains the modulation $\text{e}^{-\text{j} \pi k}$ and the complex conjugate of the channel coefficient from the image subcarrier $c^*_{-k,m}$. Again, the modulation shifts this term along the range axis. The complex conjugation of the channel coefficient inverts the velocity in the \ac{rvm}, causing the ghost object at the location of the green square in Fig.~\ref{fig_SotA}. Finally, the fourth term contains $c^*_{-k,m}$, and it results in a ghost object at the location of the blue star in Fig.~\ref{fig_SotA}. 

We conclude that \ac{iqirSotA} \ac{ofdm} generates up to three ghost objects in the \ac{rvm} for every real object in case of \ac{iq} imbalance in the transmitter and in the receiver. Two of these ghost objects are located at different velocity bins. Thus, tracking these objects over several consecutive \acp{rvm} reveals that their changes in range do not correspond to the velocity. In this work, it is assumed that tracking algorithms \cite{bar1995multitarget, bar2011tracking, blackman1999design, 1263228} identify and cancel these objects within a short period of time. The ghost object at the location of the black triangle, unfortunately, propagates in the \ac{rvm} as a real object would do. Hence, tracking algorithms cannot identify it as a ghost object, and it would be falsely interpreted as a real object. The authors of \cite{IQ_Imbalance_Rad2} suggest reducing the maximum unambiguous range to $r_\text{max}/2$ such that all objects beyond this range are automatically ignored. Depending on the parametrization of \ac{iqirSotA} \ac{ofdm} and the application, this reduction of the maximum unambiguous range may be unacceptable. Moreover, in case a real object would violate the maximum unambiguous range $r_\text{max}/2$ (e.g., the black triangle in Fig.~\ref{fig_SotA} is the real object) two of the three ghost objects caused by this real object would lie within a range of $r_\text{max}/2$ (red circle and blue star) and may be falsely classified as real objects.

\section{FRIQIR OFDM Waveform}  \label{sec:IQIR_OFDM_Radar}

\ac{iqirSotA} \ac{ofdm} offers robustness against \ac{iq} imbalance, however, at the price of either a reduced maximum unambiguous range or the existence of ghost objects that cannot be identified as such by tracking algorithms. Hence, \ac{iqir} \ac{ofdm} aims at tackling those drawbacks by choosing the subcarrier symbols in a different way.


\noindent \textit{DESIGN RULE II:}
\begin{enumerate}
\item The subcarrier symbols $s_{k,m}$ for $ 0 < k < N_\text{c}/2$ are chosen arbitrarily (e.g., by loading with data to be transmitted).
\item The subcarrier symbols $s_{k,m}$ for $k \in \{ -N_\text{c}/2,\,\, 0  \}$ are chosen to be real-valued.
\item The subcarrier symbols $s_{k,m}$ for $ -N_\text{c}/2 < k < 0$ are chosen according to 
\begin{align}
s_{k,m} ={}& s^*_{-k,m}\, \text{e}^{\text{j} \pi m}. \label{equ:IQIR_OFDM_010}
\end{align}
\end{enumerate}
In contrast to \ac{iqirSotA} \ac{ofdm} \eqref{equ:IQIR_OFDM_027a}, the phase is changed along the \ac{ofdm} symbols $m$ rather than the subcarriers $k$, which not only changes the locations of the ghost objects in the \ac{rvm} but also has huge impact on the communication aspects as shown in Sec.~\ref{sec:Comm_System}. 

The exponential term $\text{e}^{\text{j} \pi m}$ in \eqref{equ:IQIR_OFDM_010} is referred to as artificial Doppler shift and it will turn out, that it shifts the ghost objects along the velocity axis. Note that \eqref{equ:IQIR_OFDM_010} holds for positive and negative values of $k$ except for $k \in \{ -N_\text{c}/2,\,\, 0  \}$, which will be tackled later in this work.
Inserting \eqref{equ:IQIR_OFDM_010} into \eqref{equ:IQIR_OFDM_007} and dividing the result by $s_{k,m}$ leads to 
 \begin{align}
z_{k,m} ={}& \alpha^\text{Rx}_{k} \, c_{k,m}\, \alpha^\text{Tx}_{k}   \nonumber \\
& + \alpha^\text{Rx}_{k} \, c_{k,m}\, \beta^\text{Tx}_{k} \, \text{e}^{-\text{j} \pi m} \nonumber \\
& + \beta^\text{Rx}_{k} \, c^*_{-k,m}\, \left(\alpha^\text{Tx}_{-k}\right)^* \,  \text{e}^{-\text{j} \pi m} \nonumber \\
& + \beta^\text{Rx}_{k} \, c^*_{-k,m}\, \left(\beta^\text{Tx}_{-k}\right)^* 
 \label{equ:IQIR_OFDM_013} 
\end{align}
for all $k$ except for $k \in \{ -N_\text{c}/2,\,\, 0  \}$. 

The result can be interpreted as follows. The first term in \eqref{equ:IQIR_OFDM_013} contains the wanted channel coefficient $c_{k,m}$ and represents the real object, indicated by the red circle in the \ac{rvm} in Fig.~\ref{fig_IQIR_OFDM}. As discussed in Sec.~\ref{sec:SOTA}, the scaling factors $\alpha^\text{Rx}_{k} \, \alpha^\text{Tx}_{k}$ may cause an increased sidelobe level along the range axis. The second and third terms in \eqref{equ:IQIR_OFDM_013} are modulated by the artificial Doppler shift $\text{e}^{-\text{j} \pi m}$. This artificial Doppler shift moves the ghost objects along the velocity axis in the \ac{rvm} to the positions of the green square and black triangle in Fig.~\ref{fig_IQIR_OFDM}. 
The fourth term contains the complex conjugated channel coefficient $c^*_{-k,m}$ from the image subcarrier. As for \ac{iqirSotA} \ac{ofdm}, this term generates a ghost object in the \ac{rvm} at the opposite velocity of the real object (blue star in Fig.~\ref{fig_IQIR_OFDM}). 

We conclude that all ghost objects generated by \ac{iqir} \ac{ofdm} appear at the same range bin as the real object but at different velocity bins. Hence, the change in range observed through a comparison of several consecutive \acp{rvm} does not correspond to the observed velocity. Hence, it will be assumed in this work that tracking algorithms are capable of identifying and eliminating all visible ghost objects. This is in contrast to \ac{iqirSotA} \ac{ofdm}, for which not all ghost objects can be eliminated by tracking algorithms. Another advantage of \ac{iqir} \ac{ofdm} over \ac{iqirSotA} \ac{ofdm} is that no reduction of the maximum unambiguous range is necessary.

%

%

\begin{table}[!t]
\renewcommand{\arraystretch}{1.3}
\caption{System parameters.}
\label{Tab:com_sys_paramters}
\centering
\begin{tabular}{|l|r|}
\hline
 Parameter & Value  \\
 \hline \hline
 Bandwidth $B$ & $1\, \text{GHz}$ \\
 \hline
 Carrier frequency $f_\text{c}$ & $77 \, \text{GHz}$ \\
 \hline
 Number of subcarriers $N_\text{c}$ & $512$ \\
 \hline
 Number of \ac{ofdm} symbols in a burst $N_\text{sym}$ & $256$ \\
 \hline
  Length of the cyclic prefix $T_\text{cp}$ & $0.5\, \mu \text{s}$ \\
 \hline
 Number of samples in the cyclic prefix $N_\text{cp}$ & $500$ \\
 \hline
 Symbol alphabet (except for $k \in \{ -N_\text{c}/2,\,\, 0  \}$) & QPSK \\
 \hline
  Window function & Chebyshev ($120\,\text{dB}$ \\
  &  sidelobe suppr.) \\
 \hline
\end{tabular}
\end{table}

\begin{figure}[!t]
\begin{center}
\begin{tikzpicture}
\begin{semilogyaxis}[compat=newest, 
width=.90\columnwidth, height = .5\columnwidth, grid, xlabel={PAPR (dB)}, 
ylabel={$P$(PAPR $>$ abscissa)}, 
legend pos=south east, 
legend cell align=left,
legend columns=1, 
xmin = 8,
xmax = 18,
ymin = 1e-6,
ymax = 1,
legend style={
at={(0.65, 0.98)},
anchor=north west,font=\tiny}
]

\addplot[line width=1pt][color=black, solid] table[x index =0, y index =1] {SimData/PAPR_StdOFDM_Nc512.dat};
\addlegendentry{{\footnotesize Standard OFDM}}

\addplot[line width=1pt][color=gray, solid, every mark/.append style={solid},mark=star,mark repeat = 5] table[x index =0, y index =1] {SimData/PAPR_SotA_Nc512.dat};
\addlegendentry{{\footnotesize IQIR OFDM}}

\addplot[line width=1pt][color=lightgray, solid] table[x index =0, y index =1] {SimData/PAPR_PropIQIR_Nc512.dat};
\addlegendentry{{\footnotesize FRIQIR OFDM}}

\end{semilogyaxis}
\end{tikzpicture}
\caption{CCDF of the PAPR for the standard OFDM, the IQIR OFDM, and the FRIQIR OFDM transmit signals.
\label{fig:PAPR} }
\end{center}
\end{figure}
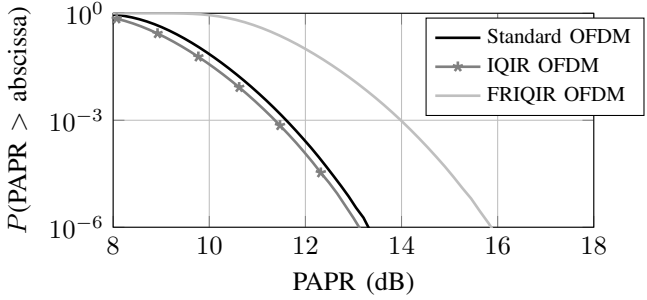

Fig.~\ref{fig:PAPR} compares the \ac{papr} \acp{ccdf} of   \ac{iqir} \ac{ofdm}  with that of standard \ac{ofdm}  and  \ac{iqirSotA} \ac{ofdm}  for the parameters listed in Tab.~\ref{Tab:com_sys_paramters}. This result indicates that for \ac{iqir} \ac{ofdm}, the \ac{papr} increases by approximately $2.4\,\text{dB}$ compared to the two others (measured at a probability of $10^{-3}$).


%

\section{Radar Simulation Results} \label{sec:Testing_SIM}

%
%
%

\ac{iqir} \ac{ofdm} is compared with \ac{iqirSotA} \ac{ofdm} and with standard \ac{ofdm} via simulations for the system parameters listed in Tab.~\ref{Tab:com_sys_paramters}.

\subsection{Simulation Setup}

The radar scenario contains $2$ objects with range $r$ and velocity $v$ chosen as $\left( 10\,\text{m}, \,\, 50 \,\text{m/s}  \right)$ and $\left( 20\,\text{m}, \,\, -20 \,\text{m/s}  \right)$ and a strong cross-talk at $\left( 0.1\,\text{m}, \,\, 0 \,\text{m/s}  \right)$. Additive measurement noise and Doppler-induced \ac{ici} were dismissed to prevent ghost objects and other effects from being hidden in the noise. 

\begin{figure}[!t]
\begin{center}
\begin{tikzpicture}
\begin{axis}[name=one, width=.70\columnwidth, height = .4\columnwidth, grid, xlabel={Subcarrier index $k$}, 
ylabel={$\epsilon^\text{Rx,FS}$ (1)},
xmin = -256,
xmax = 255,
ymin = -0.05,
ymax = 0.01,
ytick={-0.05,0},
xtick={-256, 0, 255},
]
\addplot[line width=1pt][color=black, solid] table[x index =0, y index =1] {SimData/IQ_Imbalance_LPF_Nc512.dat};
  \end{axis}

\begin{axis}[name=two, at=(one.below south), anchor=above north , width=.70\columnwidth, height = .4\columnwidth, grid, xlabel={Subcarrier index $k$},
ylabel style={align=center},  
ylabel={$\phi^\text{Rx,FS}$ (deg)},
xmin = -256,
xmax = 255,
ymin = -10,
ymax = 10,
ytick={-10,0,10},
xtick={-256, 0, 255},
]
\addplot[line width=1pt][color=black, solid] table[x index =0, y index =2] {SimData/IQ_Imbalance_LPF_Nc512.dat};
  \end{axis}

\end{tikzpicture}
\caption{FS IQ imbalance caused by the mismatch of the LPFs in the I and the Q branch of the receiver.
\label{fig:LPF_IQ} }
\end{center}
\end{figure}
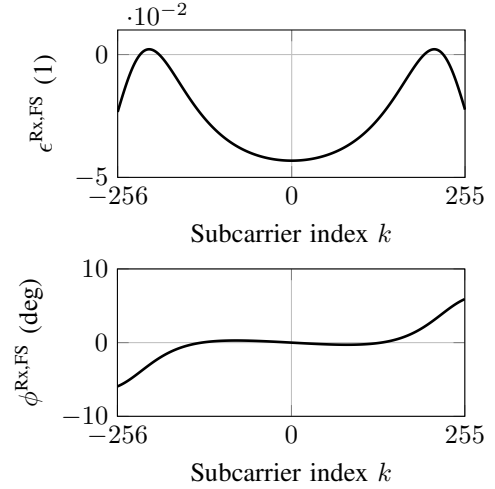

\ac{iq} imbalance sources and values are taken and adopted from \cite{IQ_Imbalance_Comm6, IQ_Imbalance_Rad1}, which is a combination of \ac{fi} and \ac{fs} components and include
\begin{itemize}
\item \textbf{FI IQ imbalance in the transmitter}: The transmitter causes \ac{fi} \ac{iq} imbalance with $\alpha^\text{Tx} = 0.9848 + \text{j}\,0.026$ and $\beta^\text{Tx} = 0.148 - \text{j}\,0.174$ (values taken from \cite{IQ_Imbalance_Rad1}), which corresponds to an amplitude imbalance of $\epsilon^\text{Tx} = 0.1503$ and a phase orthogonality mismatch of $\phi^\text{Tx} = 10^{\circ} $ according to \eqref{equ:IQIR_OFDM_004b} and \eqref{equ:IQIR_OFDM_004c}.
\item \textbf{FI IQ imbalance in the receiver}: The receiver adds \ac{fi} \ac{iq} imbalance with $\alpha^\text{Rx,FI} = 0.966 + \text{j}\,0.026$ and $\beta^\text{Rx,FI} = -0.107 + \text{j}\,0.265$, which translates to $\epsilon^\text{Rx,FI} = -0.105$ and $\phi^\text{Rx,FI} = -15^{\circ} $ (values taken from \cite{IQ_Imbalance_Rad1}). 
\item \textbf{FS IQ imbalance in the receiver}: As stated in \cite{IQ_Imbalance_Comm6}, the \acp{lpf} in the I and Q branches are a dominant source of \ac{fs} \ac{iq} imbalance. These \acp{lpf} were modeled in a similar way as in \cite{IQ_Imbalance_Comm6, Tsui_Cheby_model} by using 6th-order Chebyshev Type 1 filters, whose parametrization differs between I and Q branches. For the I branch, the ripples were set to $3\,\text{dB}$ and the passband to $0.8$ times the sampling frequency. For the Q branch, these parameters are $2\,\text{dB}$ and $0.81$, respectively. Both filters were applied at twice the sampling frequency such that all signal components are in the passband of the \acp{lpf}. The resulting \ac{fs} \ac{iq} imbalance due to the \acp{lpf} alone is shown in Fig.~\ref{fig:LPF_IQ} for each subcarrier.
\end{itemize}

\newcommand\figwidtha{0.31}
\newcommand\figwidthb{1.0}

\begin{figure*}[!t]
\centering
\begin{subfigure}{\figwidtha \textwidth}
    \includegraphics[width=\figwidthb \textwidth]{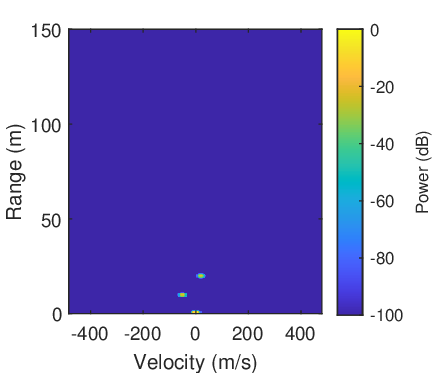}
    \caption{Std. OFDM with perfect IQ balance.}
    \label{fig:1}
\end{subfigure}
\hfill
\begin{subfigure}{\figwidtha \textwidth}
    \includegraphics[width=\figwidthb \textwidth]{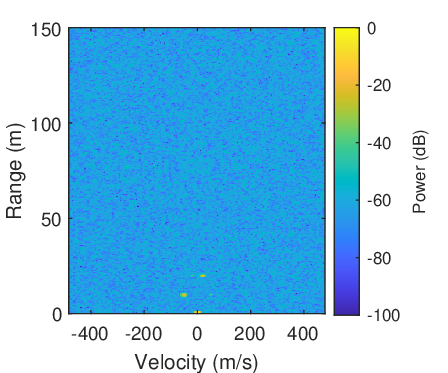}
    \caption{Std. OFDM.}
    \label{fig:2}
\end{subfigure}
\hfill
\begin{subfigure}{\figwidtha \textwidth}
    \includegraphics[width=\figwidthb \textwidth]{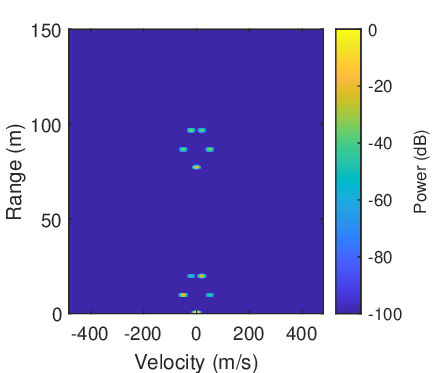}
    \caption{IQIR OFDM.}
    \label{fig:3}
\end{subfigure}
\hfill
\begin{subfigure}{\figwidtha \textwidth}
    \includegraphics[width=\figwidthb \textwidth]{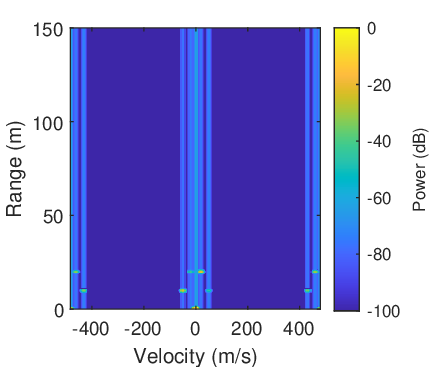}
    \caption{FRIQIR OFDM.}
    \label{fig:4}
\end{subfigure}
\hspace{1cm}
\begin{subfigure}{\figwidtha \textwidth}
    \includegraphics[width=\figwidthb \textwidth]{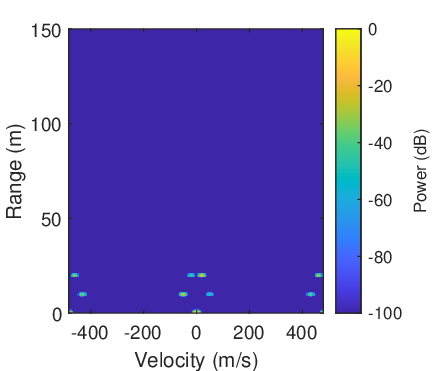}
    \caption{FRIQIR OFDM after median removal.}
    \label{fig:5}
\end{subfigure}
        
\caption{Simulated RDMs for a) a standard OFDM and perfect IQ balance. With IQ imbalance, the simulated RDMs of b)  standard OFDM, c) \ac{iqirSotA} \ac{ofdm}, d) FRIQIR OFDM as described in Sec.~\ref{sec:IQIR_OFDM_Radar}, and e) FRIQIR OFDM after removing the median values in the RDM as described in Sec.~\ref{Sec.RemMedian}.}
\label{fig_sim}
\end{figure*}

\subsection{RDM Simulation Results}

Fig.~\ref{fig_sim} shows the observed \acp{rvm}. Fig.~\ref{fig_sim} a) represents the ideal case with perfect \ac{iq} balance for standard \ac{ofdm}. 
Fig.~\ref{fig_sim} b) visualizes the increase in noise floor for standard \ac{ofdm} caused by \ac{iq} imbalance. 
Fig.~\ref{fig_sim} c) shows the \ac{rvm} for  \ac{iqirSotA} \ac{ofdm} for which ghost objects appear at the locations discussed in Sec.~\ref{sec:SOTA}. Each real object generates up to three ghost objects, of which one ghost object propagates in the \ac{rvm} like a real object and cannot be canceled by tracking algorithms. 
Fig.~\ref{fig_sim} d) shows the \ac{rvm} observed for \ac{iqir} \ac{ofdm} that shifts all ghost objects to different velocity bins, and thus, enabling tracking algorithms to cancel them. This \ac{rvm} also shows the ridges along the range axis caused by the \ac{iq} imbalance at the subcarrier indices $k \in \{ -N_\text{c}/2,\,\, 0  \}$. 

\subsection{Artefacts in the RDM and Cancellation Methods} \label{Sec.RemMedian}

Since \eqref{equ:IQIR_OFDM_013} does not hold for $k \in \{ -N_\text{c}/2,\,\, 0  \}$, these subcarriers do not incorporate ghost objects. Instead, the \ac{iq} imbalances at these subcarriers result in ridges along the range axis visible in Fig.~\ref{fig_sim} d).
The magnitude values of these ridges are much lower than the magnitude values of the ghost objects and they will be hidden in the noise floor for many practical applications. If this is not the case and the ridges exceed the noise floor, there exists a simple method for removal, which is based on the following observations. Since these ridges are caused by the \ac{iq} imbalance at the \ac{dc} subcarrier $k = 0$ and the subcarrier with the most negative frequency  $k = -N_\text{c}/2$, we observed that, when using no window function for the range \ac{dft}, these ridges consist of a \ac{dc} component and a highest-frequency component. The utilization of a window function for the range \ac{dft} smoothes these ridges such that only a \ac{dc} component remains. This \ac{dc} component is approximately $60\,\text{dB}$ below the causing object for the chosen simulation parameters and can simply be removed, e.g., by taking every vector of the \ac{rvm} corresponding to a single velocity bin and subtracting its median value. 

Fig.~\ref{fig_sim} e) shows the \ac{rvm} from  Fig.~\ref{fig_sim} d) after removing the median value from each column, which completely removes all visible ridges. This \ac{rvm} shows no increase in noise floor and all ghost objects are located at different velocity bins such that their changes in range observed over several consecutive \acp{rvm} do not align with their observed velocities.

%

\section{IQIR OFDM and FRIQIR OFDM Communication Systems} \label{sec:Comm_System}

The communication aspects of both, the proposed \ac{iqir} \ac{ofdm} and the \ac{iqirSotA} \ac{ofdm}, are analyzed in the following and compared against each other.
The investigations will start with a description of the signal processing chain and a discussion of the so-called \textit{effective channel}, which combines the true channel and the distortions caused by \ac{iq} imbalance. After that, suitable methods for channel estimation, synchronization, and data estimation will be proposed for both waveforms, that are  capable of dealing with the observed effective channel and exploiting the redundancy in the transmit data introduced. The resulting communication systems will simply be referred to as \ac{iqirSotA} \ac{ofdm} and \ac{iqir} \ac{ofdm} for the sake of brevity. Similarly, a standard \ac{ofdm} communication system is simply denoted as standard \ac{ofdm}.

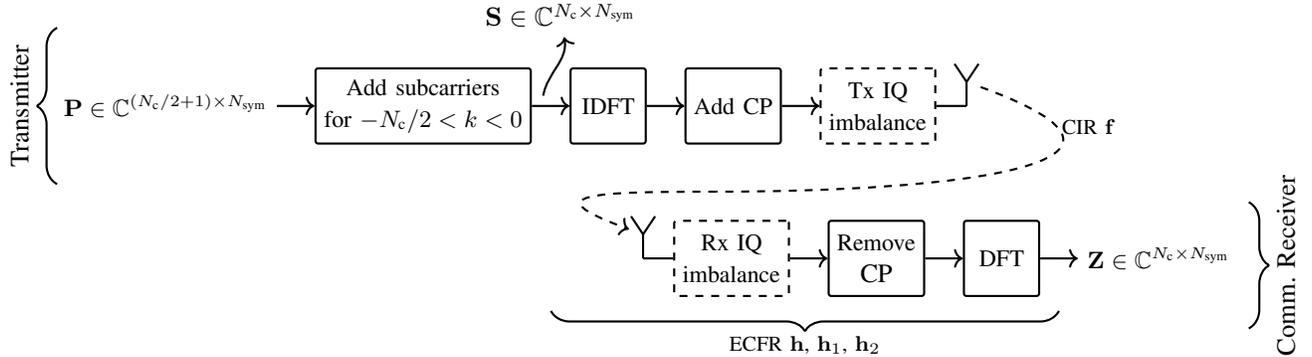
\begin{figure*}[!t]
\centering
\begin{tikzpicture}[scale=0.8, style=thick, rounded corners=1pt,inner sep=3.2pt,node distance=.8cm,every text node part/.style={align=center}]

\node[ minimum height = 1cm, minimum width = 1cm] (state0c){\small $\m{P} \in \mathbb{C}^{(N_\text{c}/2+1) \times N_\text{sym}}$};

\node[draw,right=0.5cm of state0c, minimum height = 1cm, minimum width = 1cm](state0){\small Add subcarriers \\ \small for $-N_\text{c}/2 < k < 0$};

\draw[->] (state0c.east)  -- (state0.west);

\node[right=0.0cm of state0](yarr1){};
\node[above right =0.8cm and -1.0cm of yarr1](yarr2){$\m{S}\in \mathbb{C}^{N_\text{c} \times N_\text{sym}}$};
\draw [->, shorten <=0.01cm, shorten >=0.01cm, style=solid] (yarr1) to [out=80,in=280] (yarr2);

\node[draw,right=0.5cm of state0, minimum height = 1cm, minimum width = 1cm](state1){\small IDFT};

\draw[->] (state0.east)  -- (state1.west);

\node[draw,right=0.5cm of state1, minimum height = 1cm, minimum width = 1cm](state2){\small Add CP};

\draw[->] (state1.east)  -- (state2.west);

\node[draw,dashed,right=0.5cm of state2, minimum height = 1cm, minimum width = 1cm](state2x){\small Tx IQ \\ \small imbalance};

\draw[->] (state2.east)  -- (state2x.west);

\draw (state2x.east)  -- ++(0.5,0) coordinate (Ant1u);
\draw (Ant1u)  -- ++(0,0.4) coordinate (IntAnt1u);
\draw (IntAnt1u)  -- ++(-0.2,0.3);
\draw (IntAnt1u)  -- ++(0.2,0.3);

\node[draw,dashed,below right =1.0cm and -3.5cm of state2x, minimum height = 1cm, minimum width = 1cm](state6){\small Rx IQ \\ \small imbalance};

\draw (state6.west)  -- ++(-0.5,0) coordinate (Ant2);
\draw (Ant2)  -- ++(0,0.4) coordinate (IntAnt2);
\draw (IntAnt2)  -- ++(-0.2,0.3);
\draw (IntAnt2)  -- ++(0.2,0.3);

\node[below right =0.5cm and -1cm of state2, minimum height = 0cm, minimum width = 0cm](IntAnt1below){};

\draw [-, shorten <=0.2cm, style=dashed, out=-20,in=0, distance = 4cm] (IntAnt1u) to  (IntAnt1below.center);

\draw [->, shorten >=0.2cm, style=dashed, out=180,in=160, distance = 2cm] (IntAnt1below.center) to  (IntAnt2);


\node[draw,right=0.5cm of state6, minimum height = 1cm, minimum width = 1cm](state6e){\small Remove \\ CP};


\draw[->] (state6.east)  -- (state6e.west);

\node[draw,right=0.5cm of state6e, minimum height = 1cm, minimum width = 1cm](state7){\small DFT};


\draw[->] (state6e.east)  -- (state7.west);

\node[ right=0.5cm of state7, minimum height = 1cm, minimum width = 1cm] (state8){\small $\m{Z} \in \mathbb{C}^{N_\text{c} \times N_\text{sym}}$};

\draw[->] (state7.east)  -- (state8.west);

%

\node[below right =0.05cm and 3.6cm of state2.east, minimum height = 0cm, minimum width = 0cm](cb1){\footnotesize CIR $\ve{f}$};

\node[below right =2.6cm and 0.0cm of state0.east, minimum height = 0cm, minimum width = 0cm](cb1){};

\draw [decorate,decoration={brace,mirror, amplitude=8pt},xshift=0pt,yshift=0pt]
(cb1) -- ++(8.6,0.0) node [black,midway,yshift=-0.45cm] 
{\footnotesize ECFR  $\ve{h}$, $\ve{h}_1$, $\ve{h}_2$};

\node[align=center] (A1) at (-1.8,-1.3) {};

	\draw [decorate,decoration={brace, amplitude=8pt},xshift=0pt,yshift=0pt]
	(A1.center) -- ++(0,2.6) node [black,midway,xshift=-0.5cm,rotate=90] 
	{Transmitter};
	
	\node[align=center] (A2) at (18.0,-1.6) {};

	\draw [decorate,decoration={brace, amplitude=8pt},xshift=0pt,yshift=0pt,rotate=180]
	(A2.center) -- ++(0,2.1) node [black,midway,xshift=0.5cm,rotate=90] 
	{Comm. Receiver};
	
\end{tikzpicture}
\caption{Visualization of basic signal processing blocks in the transmitter and the communication receiver, the CIR, and the ECFR including its covered processing blocks. The analog front-end and the parallel-to-serial conversion are not shown for simplicity.}
\label{fig:Comm_setup_ECIR}
\end{figure*}

\subsection{Signal Processing Chain} \label{sec:Signal_chain}

The transmitter's and receiver's basic signal processing blocks and the channel are visualized in Fig.~\ref{fig:Comm_setup_ECIR}, which is valid for \ac{iqirSotA} \ac{ofdm} and \ac{iqir} \ac{ofdm}. The rows of the matrix $\m{P} \in \mathbb{C}^{(N_\text{c}/2+1) \times N_\text{sym}}$ contain the subcarrier symbols for $ 0 < k < N_\text{c}/2$ as well as the two real-valued subcarrier symbols for $k \in \{ -N_\text{c}/2,\,\, 0  \}$  for all $N_\text{sym}$ \ac{ofdm} symbols. The first processing block adds the remaining subcarriers $ -N_\text{c}/2 < k < 0$ according to the design rule I or II, yielding the matrix $\m{S} \in \mathbb{C}^{N_\text{c} \times N_\text{sym}}$. 
Each row of $\m{S}$ represents a subcarrier, and each column of $\m{S}$ represents an \ac{ofdm} symbol in frequency domain. Then, the \ac{ofdm} symbols are transformed into time domain via an \ac{idft} and extended by a \ac{cp}. After applying \ac{iq} imbalance in the same way as for the simulations in Sec.~\ref{sec:Testing_SIM}, the time-domain signal is convolved with the \ac{cir} $\ve{f} \in \mathbb{C}^{N_\ve{f}}$ of an assumed length $N_\ve{f} \leq N_\text{c}$. The receiver adds additional \ac{iq} imbalance as described in Sec.~\ref{sec:Testing_SIM}, removes the \ac{cp}, and performs a \ac{dft}.
The resulting \ac{ofdm} symbols are then stored in the matrix $\m{Z} \in \mathbb{C}^{N_\text{c} \times N_\text{sym}}$.


\subsection{Channel Model} \label{sec:Channel_Model}

In this section we model channel models for IQIR and fRIQIR OFDM, which contains additional deterministic and random components, while for standard OFDM all components are random

In this section, we develop channel models that include \ac{iq} imbalance for all considered waveforms and that will be used later in this work for the development of  channel estimation, synchronization, and data estimation methods. 


The time domain representation of the channel between the transmit and receive antennas is the \ac{cir} $\ve{f}$. This \ac{cir} $\ve{f}$ would also be the outcome of a channel estimation procedure in the ideal case of perfect \ac{iq} balance and no additive measurement noise. However, in case of \ac{iq} imbalance, a different channel is observed. This channel is referred to as \textit{effective channel} and it combines the \ac{cir} $\ve{f}$ and the distortions caused by \ac{iq} imbalance. 
This effective channel in time domain is referred to as \ac{ecir} $\ve{g} \in \mathbb{C}^{N_\text{c}}$, and in frequency domain as \ac{ecfr} $\ve{h} \in \mathbb{C}^{N_\text{c}}$ given by 
\begin{align}
	\ve{h} &={} \m{F}_{N_\text{c}} \ve{g} \in \mathbb{C}^{N_\text{c}}, \label{equ:IQIR_OFDM_Comm001} 
\end{align}
with $ \m{F}_{N_\text{c}}$ denoting the \ac{dft} matrix of size $N_\text{c} \times N_\text{c}$.

In the following, the effective channels for the considered waveforms are analyzed. This analysis will focus on the channel's time domain representation \ac{ecir} $\ve{g}$ that is observed after applying an \ac{idft} on the result of an element-wise division of a received frequency-domain \ac{ofdm} symbol $\ve{y}\in \mathbb{C}^{N_\text{c}}$ by the transmitted frequency-domain \ac{ofdm} symbol $\ve{s}\in \mathbb{C}^{N_\text{c}}$  according to $\ve{g} = \m{F}_{N_\text{c}}^{-1} \left( \ve{y} \oslash \ve{s} \right) $.
Here, $\ve{y}$ does not contain any additive measurement noise. It will turn out that for standard \ac{ofdm}, \ac{iq} imbalances cause random distortions in the \ac{ecir} $\ve{g}$ due to the random nature of the subcarrier symbols. However, for \ac{iqirSotA} \ac{ofdm} and for \ac{iqir} \ac{ofdm}, \ac{iq} imbalances produce deterministic components in the \ac{ecir} $\ve{g}$ discussed in the following.


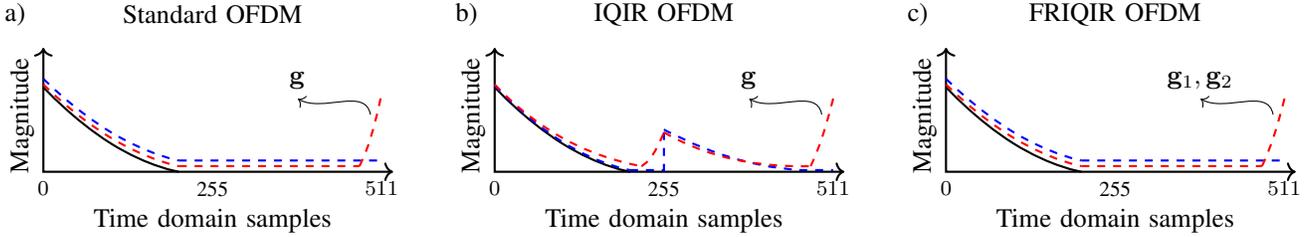
\begin{figure*}[!t]
\begin{tikzpicture}[scale=0.75]

\def\d{8}

\node at (-0.5,2.8) {a) };
\node at (3,2.8) {Standard OFDM};

   \draw [<->,thick] (0,2.2) node (yaxis) [above] {}
        |- (6.2,0) node (xaxis) [below] {};    
    \node[below] at (0,0 ) {\footnotesize $0$}; 
    \node[below] at (6,0 ) {\footnotesize $511$}; 
    \node[left, anchor=south, rotate=90] at (0,1) {Magnitude};
    \node[below] at (3,0) {\footnotesize 255};
    \node[below] at (3,-0.5) {Time domain samples};

	
\draw [black,thick, xshift=0cm] plot [smooth, tension=1] coordinates { (0,1.5) (1.2,0.5) (2.4,0) };

\draw [blue,thick, dashed, xshift=0cm] plot [smooth, tension=1] coordinates { (0,1.65) (1.2,0.7) (2.4,0.2) };
\draw [blue,thick, dashed, xshift=0cm] plot [smooth, tension=1] coordinates { (2.4,0.2) (6,0.2) };

\draw [red,thick, dashed, xshift=0cm] plot [smooth, tension=1] coordinates { (0,1.55) (1.2,0.6) (2.4,0.1) };
\draw [red,thick, dashed, xshift=0cm] plot [smooth, tension=1] coordinates { (2.4,0.1)  (5.6,0.1) };
\draw [red,thick, dashed, xshift=0cm] plot [smooth, tension=1] coordinates { (5.6,0.1) (5.8,0.6) (6,1.4) };

\node[above] at (4.5, 1.3) {$\ve{g}$};
\draw [->, shorten <=0.01cm, shorten >=0.01cm, style=solid] (5.8, 1) to [out=100,in=-30] (4.5, 1.3);

\begin{scope}[shift={(\d,0)}]  

\node at (-0.5,2.8) {b) };
\node at (3,2.8) {IQIR OFDM};

	    \draw [<->,thick] (0,2.2) node (yaxis) [above] {}
        |- (6.2,0) node (xaxis) [below] {};    
    \node[below] at (0,0 ) {\footnotesize $0$}; 
    \node[below] at (6,0 ) {\footnotesize $511$}; 
    \node[left, anchor=south, rotate=90] at (0,1) {Magnitude};
    \node[below] at (3,0) {\footnotesize 255};
    \node[below] at (3,-0.5) {Time domain samples};

	
\draw [black,thick, xshift=0cm] plot [smooth, tension=1] coordinates { (0,1.5) (1.2,0.5) (2.4,0) };

\draw [blue,dashed,thick, xshift=0cm] plot [smooth, tension=1] coordinates { (0,1.53) (1.2,0.53) (2.4,0.03) };
\draw [blue,dashed,thick, xshift=0cm] plot [smooth, tension=1] coordinates { (2.4,0.03)  (3,0.03) };
\draw [blue,dashed,thick, xshift=0cm] plot coordinates { (3,0) (3,0.75) };
\draw [blue,dashed,thick, xshift=0cm] plot [smooth, tension=1] coordinates { (3,0.75) (4.2,0.25) (5.4,0.03) };
\draw [blue,dashed,thick, xshift=0cm] plot [smooth, tension=1] coordinates { (5.4,0.03)  (6,0.03) };

\draw [red,thick, dashed, xshift=0cm] plot [smooth, tension=1] coordinates { (0,1.55) (1.2,0.6) (2.6,0.1) };
\draw [red,thick, dashed, xshift=0cm] plot [smooth, tension=1] coordinates { (2.6,0.1) (2.8,0.3) (3,0.7) };
\draw [red,thick, dashed, xshift=3cm] plot [smooth, tension=1] coordinates { (0,0.7) (1.2,0.23) (2.6,0.1) };
\draw [red,thick, dashed, xshift=3cm] plot [smooth, tension=1] coordinates { (2.6,0.1) (2.8,0.6) (3,1.3) };

\node[above] at (4.5, 1.3) {$\ve{g}$};
\draw [->, shorten <=0.01cm, shorten >=0.01cm, style=solid] (5.8, 1) to [out=100,in=-30] (4.5, 1.3);

 \end{scope}

  \begin{scope}[shift={(2*\d,0)}]  
  
  \node at (-0.5,2.8) {c) };
\node at (3,2.8) {FRIQIR OFDM};

	   \draw [<->,thick] (0,2.2) node (yaxis) [above] {}
        |- (6.2,0) node (xaxis) [below] {};    
    \node[below] at (0,0 ) {\footnotesize $0$}; 
    \node[below] at (6,0 ) {\footnotesize $511$}; 
    \node[left, anchor=south, rotate=90] at (0,1) {Magnitude};
    \node[below] at (3,0) {\footnotesize 255};
    \node[below] at (3,-0.5) {Time domain samples};

	
\draw [black,thick, xshift=0cm] plot [smooth, tension=1] coordinates { (0,1.5) (1.2,0.5) (2.4,0) };

\draw [blue,thick, dashed, xshift=0cm] plot [smooth, tension=1] coordinates { (0,1.65) (1.2,0.7) (2.4,0.2) };
\draw [blue,thick, dashed, xshift=0cm] plot [smooth, tension=1] coordinates { (2.4,0.2) (6,0.2) };

\draw [red,thick, dashed, xshift=0cm] plot [smooth, tension=1] coordinates { (0,1.55) (1.2,0.6) (2.4,0.1) };
\draw [red,thick, dashed, xshift=0cm] plot [smooth, tension=1] coordinates { (2.4,0.1)  (5.6,0.1) };
\draw [red,thick, dashed, xshift=0cm] plot [smooth, tension=1] coordinates { (5.6,0.1) (5.8,0.6) (6,1.3) };

\node[above] at (4.5, 1.3) {$\ve{g}_1, \ve{g}_2$};
\draw [->, shorten <=0.01cm, shorten >=0.01cm, style=solid] (5.8, 1) to [out=100,in=-30] (4.5, 1.3);

 \end{scope}

\end{tikzpicture}
\caption{Schematic visualization of the components of the ECIR $\ve{g}$ with $N_\text{c}=512$ for a) standard OFDM,  b) IQIR OFDM, and c) FRIQIR OFDM. The black solid line sketches a CIR, the blue dashed line the CIR and the effects caused by FI IQ imbalance, and the red dashed line indicates the CIR plus the effects caused by FI and FD IQ imbalance.}
\label{fig_All_ECIRs}
\end{figure*}

\vspace{1mm} 

\noindent \textbf{Effective channel for standard OFDM}: 
Fig.~\ref{fig_All_ECIRs} a) describes the observed \ac{ecir} $\ve{g}$  for standard \ac{ofdm}, whose main component is the \ac{cir} $\ve{f}$ (black, solid). \ac{fi} \ac{iq} imbalance causes signal components from its image to leak into the subcarrier (cf. \eqref{equ:IQIR_OFDM_005}). Since the subcarrier symbols are all random, this leakage acts as an additional source  of noise. Instead of modeling it as such, one can also model it as random distortions added to all time domain samples of the \ac{ecir} $\ve{g}$ (blue, dashed). \ac{fs} \ac{iq} imbalance adds a different kind of distortion. According to \eqref{equ:IQIR_OFDM_005}, \ac{fs} \ac{iq} imbalance can be modeled as multiplications of the subcarrier symbols with $\alpha^\text{Tx}_{k}$ and $\beta^\text{Tx}_{k}$ that vary over $k$. This multiplication in frequency domain translates to a circular convolution in time domain that spreads channel components along the time axis. The resulting \ac{ecir} $\ve{g}$ after this circular convolution is sketched in Fig.~\ref{fig_All_ECIRs} a) (red, dashed). 

Note that the \ac{ecir} $\ve{g}$ for standard OFDM contains parts with meaningful channel information and parts occupied by random distortions only. For instance, for the channels and the parametrization used in Sec.~\ref{sec:BER_performance}, simulations showed that elements of $\ve{g}$ with indices from $257$ to $412$ contain random distortions only.

\vspace{1mm} 

\noindent \textbf{Effective channel for \ac{iqirSotA} \ac{ofdm}}: 
Fig.~\ref{fig_All_ECIRs} b) schematically sketches the observed \ac{ecir} $\ve{g}$ and its components. The first component is the \ac{cir} $\ve{f}$ (black, solid).  For \ac{iqirSotA} \ac{ofdm}, \ac{fi} \ac{iq} imbalance does not add random distortions as observed for standard \ac{ofdm}, since the subcarrier symbols are not chosen randomly. Instead, 
\ac{fi} \ac{iq} imbalance adds another component  (blue, dashed), which is a scaled and shifted version of the \ac{cir}. The origin of this component is \eqref{equ:IQIR_OFDM_027a} in design rule I that contains the term $\text{e}^{\text{j} \pi k}$. This term translates to a circular shift in time domain by $N_\text{c}/2$ samples as indicated in Fig.~\ref{fig_All_ECIRs} b). Also for this communication system, \ac{fs} \ac{iq} imbalance causes a circular convolution in time domain, which spreads the two mentioned components along the time axis resulting in the final \ac{ecir} (red, dashed). 

Note that the \ac{ecir} $\ve{g}$ for \ac{iqirSotA} \ac{ofdm} contains meaningful channel information at all elements such that there are no parts occupied by random distortions only.

\vspace{1mm} 

\noindent \textbf{Effective channel for FRIQIR OFDM}: 
Fig.~\ref{fig_All_ECIRs} c) sketches the \ac{ecir} for \ac{iqir} \ac{ofdm}, whose main component is the \ac{cir} itself (black, solid). \ac{fi} \ac{iq} imbalance causes distortions, since signal components from its image leak into the subcarriers. These distortions are not random, since the subcarrier symbols are not chosen randomly, however, these distortions change their signs from \ac{ofdm} symbol to \ac{ofdm} symbol due to the term $\text{e}^{\text{j} \pi m}$ in design rule II. As a result, we observe two different deterministic \acp{ecir}, one for even and one for odd values of the \ac{ofdm} symbol index $m$. For both \acp{ecir}, \ac{fs} \ac{iq} imbalance causes a circular convolution in time domain. The two final \acp{ecir} will be referred to as $\ve{g}_1$ and $\ve{g}_2$ for even and odd values of $m$, respectively (red, dashed). 

Similar to standard \ac{ofdm}, only parts of the \acp{ecir} $\ve{g}_1$ and $\ve{g}_2$ contain meaningful channel information, while the parts with indices from $257$ to $412$ contain distortions only.

\subsection{Signal Model} \label{sec:Signal_Model}

%

In the following, a model that connects the transmitted \ac{ofdm} symbols $\m{S}$ with the received \ac{ofdm} symbols  $\m{Z}$ in frequency domain is presented. Although the notation used in this model considers only a single \ac{ecir} $\ve{g}$ and thus, only applies for the standard \ac{ofdm} and the \ac{iqirSotA} \ac{ofdm}, it can easily be extended to \ac{iqir} \ac{ofdm} when replacing $\ve{g}$ by $\ve{g}_1$ and $\ve{g}_2$ for even and odd values of $m$, respectively. 

Let $ \varphi_{m}\in \mathbb{R}$ be a \ac{cpe} that models phase rotations for the $m$th received \ac{ofdm} symbol, e.g., due to a relative velocity between transmitter and receiver. Moreover, let $ \m{H}= \text{diag}\left( \ve{h} \right) =  \text{diag}\left( \m{F}_{N_\text{c}} \ve{g} \right) \in \mathbb{C}^{N_\text{c} \times N_\text{c}}$, be referred to as channel matrix. Then, the connection between the transmitted \ac{ofdm} symbols $\m{S}$ and the received \ac{ofdm} symbols $\m{Z}$  in frequency domain is modeled as 
\begin{align}
	\m{Z} &={} \m{H}  \m{S} \m{\Lambda} + \m{N},\label{equ:modelMIMO_OFDM_011}
\end{align}
where $\m{N} \in \mathbb{C}^{N_\text{c} \times N_\text{sym}}$ contains uncorrelated zero-mean white Gaussian noise samples with variance $N_\text{c} \sigma_\text{n}^2$ and with $\sigma_\text{n}^2$ representing the time domain noise variance. The elements of the diagonal matrix $\m{\Lambda} \in \mathbb{C}^{N_\text{sym} \times N_\text{sym}}$ model the \acp{cpe} and are given by  $\text{e}^{\text{j} \varphi_{m}}$. 

Let the $m$th columns of $\m{S}$, $\m{Z}$, and $\m{N}$ be  $\ve{s}_{m}\in \mathbb{C}^{N_\text{c}}$, $\ve{z}_{m}\in \mathbb{C}^{N_\text{c}}$, and $\ve{n}_{m}\in \mathbb{C}^{N_\text{c}}$, respectively, then \eqref{equ:modelMIMO_OFDM_011} translates to 
\begin{align}
	\ve{z}_{m} &={}   \m{H}  \ve{s}_{m} \text{e}^{\mathrm{j} \varphi_m}  + \ve{n}_{m}. \label{equ:IEEE_RDM_data001} 
\end{align}

\subsection{Channel Estimation} \label{sec:Channel_Estimation}



We will first revise a well-known channel estimation procedure for standard \ac{ofdm} and adapt it to account for \ac{fs} \ac{iq} imbalance. The description of this estimation procedure serves then as a basis for describing the channel estimation procedures for \ac{iqirSotA} \ac{ofdm} and \ac{iqir} \ac{ofdm}.

\vspace{1mm} 

\noindent \textbf{Channel estimation for standard OFDM}:
To enable channel estimation, we assume that a preamble consisting of $N_\text{pr}$ repetitions of a well defined preamble \ac{ofdm} symbol is preceding the data part of a burst. 
The frequency domain representation of this preamble \ac{ofdm} symbol is $\ve{s}_\text{pr}$ and its elements all have unit magnitude values. We will further assume that $N_\text{pr}$ is an even number.
Equation \eqref{equ:IEEE_RDM_data001} serves as a basis for channel estimation. Exchanging the roles of the channel and the transmit symbols, and incorporating \eqref{equ:IQIR_OFDM_Comm001}, yields 
\begin{align}
	\ve{z}_{m} &={}  \text{diag} \left( \ve{s}_\text{pr} \right) \m{F}_{N_\text{c}} \ve{g} \text{e}^{\mathrm{j} \varphi_m} + \ve{n}_{m}. \label{equ:IEEE_RDM_prea001} 
\end{align}
Next, the $N_\text{pr}$ received preamble \ac{ofdm} symbols are synchronized by means of reversing the \acp{cpe}. As in \cite{Lang_RDM_JP}, we estimate the \ac{cpe} of the $m$th received preamble \ac{ofdm} symbol \ac{wrt} the first preamble \ac{ofdm} symbol via \cite{classen1994frequency, Diss_Hofbauer, huemer2002simulation} 
\begin{align}
\widehat{\varphi}_m = \text{arg} \left( \ve{z}_{0}^H \ve{z}_{m} \right)	 \label{equ:IEEE_RDM_pilot006a} 
\end{align}
for $0 \leq m < N_\text{pr}$.
These estimates are used for reversing the \acp{cpe} of all received preamble \ac{ofdm} symbols according to
\begin{align}
\vef{z}_{m} &={}  \ve{z}_{m} \, \text{e}^{-\mathrm{j} \widehat{\varphi}_m}.	 \label{equ:IEEE_RDM_pilot007h} 
\end{align}
These synchronized preamble \ac{ofdm} symbols are then averaged according to \cite{Lang_Asilomar_2014, Lang_RDM_JP}
\begin{align}
	\bar{\vef{z}}  &={}  \frac{1}{N_\text{pr}} \sum_{m=0}^{N_\text{pr}-1} \vef{z}_{m}. \label{equ:IEEE_RDM_CIR032lk} 
\end{align}
With 
\begin{align}
	\bar{\ve{n}}  &={}  \frac{1}{N_\text{pr}} \sum_{m=0}^{N_\text{pr}-1} \ve{n}_{m}\, \text{e}^{-\mathrm{j} \widehat{\varphi}_m} \label{equ:IEEE_RDM_CIR0xyz} 
\end{align}
and with the assumption of sufficiently accurate synchronization, \eqref{equ:IEEE_RDM_prea001} simplifies to
\begin{align}
	\bar{\vef{z}}  &\approx{} \text{diag} \left( \ve{s}_\text{pr} \right) \m{F}_{N_\text{c}} \ve{g} +\bar{\ve{n}}, \label{equ:IEEE_RDM_CIR032ll} 
\end{align}
Applying the \ac{blue} \cite{Kay-Est.Theory, Diss_Lang_Oliver,  salehi2007digital, Lang_Asilomar_2014} on \eqref{equ:IEEE_RDM_CIR032ll}  and incorporating the knowledge that the \ac{ecir} $\ve{g}$ does not contain meaningful channel information on all of its elements (cf. discussion in Sec.~\ref{sec:Channel_Model}) yields the final estimate $\veh{g} \in \mathbb{C}^{N_\text{c}}$ of the \ac{ecir}. Due to the assumption of unit magnitude values of the elements in $\ve{s}_\text{pr}$, the estimate $\veh{g}$ is simply given by 
\begin{align}
	\veh{g}  &={} \m{M}_\ve{g} \m{F}_{N_\text{c}}^{-1} \left(\text{diag} \left( \ve{s}_\text{pr} \right) \right)^{-1} \bar{\vef{z}}, \label{equ:IEEE_RDM_CIR032lm} 
\end{align}
where $\m{M}_\ve{g} \in \mathbb{R}^{N_\text{c} \times N_\text{c}}$ is a diagonal matrix that sets the elements with indices from $257$ to $412$ to zero and that leaves the remaining elements unaltered. Based on \eqref{equ:IEEE_RDM_CIR032lm}, the \ac{blue} for the channel matrix $\m{H}$ follows as
\begin{align}
	 \mh{H}&={}   \text{diag}\left( \veh{h} \right) =   \text{diag}\left( \m{F}_{N_\text{c}} \veh{g} \right). \label{equ:IEEE_RDM_data002} 
\end{align}

\vspace{1mm} 

\noindent \textbf{Channel estimation for \ac{iqirSotA} \ac{ofdm}}: 
The channel estimation procedure for this communication system is also based on $N_\text{pr}$ repetitions of a well defined frequency domain preamble \ac{ofdm} symbol\footnote{The roman number I indicates design rule I.} $\ve{s}_\text{pr}^{\text{I}}$ preceding the data part of a burst. 
It is assumed that $\ve{s}_\text{pr}^{\text{I}}$ fulfills design rule I and that all elements of $\ve{s}_\text{pr}^{\text{I}}$ have unit magnitude values. The first processing steps up to the model in \eqref{equ:IEEE_RDM_CIR032ll} follow the very same steps as for standard \ac{ofdm}. The \ac{blue} for the \ac{ecir} $\ve{g}$ differs from that for standard \ac{ofdm}, since $\ve{g}$ contains meaningful channel information on all samples (cf. discussion in Sec.~\ref{sec:Channel_Model}) leading to 
\begin{align}
	\veh{g}  &={} \m{F}_{N_\text{c}}^{-1} \left(\text{diag} \left( \ve{s}_\text{pr}^{\text{I}} \right) \right)^{-1} \bar{\vef{z}} \in \mathbb{C}^{N_\text{c}}. \label{equ:IEEE_RDM_CIR032lmd} 
\end{align}
The final estimate  $\mh{H}$ is again obtained via \eqref{equ:IEEE_RDM_data002}. 

\vspace{1mm}

\noindent \textbf{Channel estimation for FRIQIR OFDM}:
Here, we utilize two constant frequency domain preamble \ac{ofdm} symbols, namely $\ve{s}_\text{pr,1}^{\text{II}}$ transmitted during the first $N_\text{pr}/2$ even \ac{ofdm} symbols, and $\ve{s}_\text{pr,2}^{\text{II}}$ transmitted during the first $N_\text{pr}/2$ odd \ac{ofdm} symbols. It is assumed that $\ve{s}_\text{pr,1}^{\text{II}}$ and $\ve{s}_\text{pr,2}^{\text{II}}$ are designed to fulfill design rule II and that their elements have unit magnitude values.

The $N_\text{pr}/2$ received even and odd preamble \ac{ofdm} symbols are synchronized separately to undo the phase rotations caused by the \acp{cpe}. These \acp{cpe} are estimated via
\begin{align}
\widehat{\varphi}_m = \text{arg} \left( \ve{z}_{0}^H \ve{z}_{m} \right)	 \label{equ:IEEE_RDM_pilot006a1} 
\end{align}
for $m = 0, 2, 4, \hdots, N_\text{pr}-2$, and via 
\begin{align}
\widehat{\varphi}_m = \text{arg} \left( \ve{z}_{1}^H \ve{z}_{m} \right)	 \label{equ:IEEE_RDM_pilot006a2} 
\end{align}
for $m = 1, 3, 5, \hdots, N_\text{pr}-1$.
These estimated \acp{cpe} are used to synchronize the received preamble \ac{ofdm} symbols according to \eqref{equ:IEEE_RDM_pilot007h}, yielding the synchronized preamble \ac{ofdm} symbols $\vef{z}_{m}$. 
These are then averaged separately for even and odd values of $m$ according to
\begin{align}
	\bar{\vef{z}}_1  &={}  \frac{1}{N_\text{pr}/2} \sum_{m=0}^{N_\text{pr}/2-1} \vef{z}_{2m}, \label{equ:IEEE_RDM_CIR032lk1}  \\
	\bar{\vef{z}}_2  &={}  \frac{1}{N_\text{pr}/2} \sum_{m=0}^{N_\text{pr}/2-1} \vef{z}_{2m+1}. \label{equ:IEEE_RDM_CIR032lk2} 
\end{align}
With that, the \acp{blue} for the \acp{ecir} $\ve{g}_1$ and $\ve{g}_2$ are given by 
\begin{align}
	\veh{g}_1  &={} \m{M}_\ve{g} \m{F}_{N_\text{c}}^{-1} \left(\text{diag} \left( \ve{s}_\text{pr,1}^{\text{II}} \right) \right)^{-1} \bar{\vef{z}}_1 \in \mathbb{C}^{N_\text{c}}, \label{equ:IEEE_RDM_CIR032lm1} \\
	\veh{g}_2  &={} \m{M}_\ve{g} \m{F}_{N_\text{c}}^{-1} \left(\text{diag} \left( \ve{s}_\text{pr,2}^{\text{II}} \right) \right)^{-1} \bar{\vef{z}}_2 \in \mathbb{C}^{N_\text{c}}, \label{equ:IEEE_RDM_CIR032lm2} 
\end{align}
where $\m{M}_\ve{g} \in \mathbb{R}^{N_\text{c} \times N_\text{c}}$ is a diagonal matrix that sets the elements with indices from $257$ to $412$ to zero and that leaves the remaining elements unaltered. Finally, the \acp{blue} for the two channel matrices $\mh{H}_1$ and $\mh{H}_2$ for even and odd values of $m$ are given by 
\begin{align}
	 \mh{H}_1 &={}   \text{diag}\left( \veh{h}_1 \right) =   \text{diag}\left( \m{F}_{N_\text{c}} \veh{g}_1 \right)  \label{equ:IEEE_RDM_data0021} \\
	 \mh{H}_2 &={}   \text{diag}\left( \veh{h}_2 \right) =   \text{diag}\left( \m{F}_{N_\text{c}} \veh{g}_2 \right).  \label{equ:IEEE_RDM_data0022} 
\end{align}


\subsection{Synchronization} \label{sec:Synchronization}

Synchronization of the \ac{ofdm} symbols by means of reversing the \acp{cpe} for $m \geq N_\text{pr}$ is based on $N_\text{p}$ pilot subcarriers whose subcarrier symbols are known to the receiver. Hence, every transmitted \ac{ofdm} symbol $\ve{s}_{m}$ for $m \geq N_\text{pr}$ contains $N_\text{p}$ pilot subcarriers and $N_\text{d} = N_\text{c} - N_\text{p}$ data subcarriers.
It is assumed that $N_\text{d}$ and $N_\text{p}$ are even numbers. The subcarriers $k \in \{ -N_\text{c}/2,\,\, 0  \}$ are chosen to be pilot subcarriers as their subcarrier symbols can only be real-valued according to both design rules and thus, have limited suitability for carrying data. Other pilot subcarriers are chosen pair-wise from positive and negative subcarrier indices $\{ -k,\,\, k  \}$ to fulfill the design rules. Based on that, the synchronization procedures for the considered communication systems are described in the following.


\vspace{1mm} 

\noindent \textbf{Synchronization for standard OFDM}:
Here, the synchronization procedure is not influenced by \ac{iq} imbalance as  was the case for channel estimation. Hence, well-known synchronization procedures based on the estimated pilot subcarrier symbols at the output of the \ac{lmmse} data estimator can be used to reverse the \acp{cpe}. We refer to existing literature for more details \cite{Kay-Est.Theory, huemer2011classical, huemer2002simulation, classen1994frequency, Diss_Hofbauer, Lang_RDM_JP, Van_Nee_OFDM, salehi2007digital, haykin2013digital}.

\vspace{1mm} 

\noindent \textbf{Synchronization for \ac{iqirSotA} \ac{ofdm}}: 
Here, synchronization based on the estimated pilot subcarrier symbols at the output of the \ac{lmmse} data estimator is not possible due to a processing step discussed later in this work. Instead, a separate \ac{lmmse} estimator for estimating the pilot subcarrier symbols is employed.

Let $m \geq N_\text{pr}$  and $\ve{s}_{m}^\text{p}\in \mathbb{C}^{N_\text{p}}$, $\ve{z}_{m}^\text{p}\in \mathbb{C}^{N_\text{p}}$, and $\ve{n}_{m}^\text{p}\in \mathbb{C}^{N_\text{p}}$  be the sub-vectors of their respective signal vectors $\ve{s}_{m}$, $\ve{z}_{m}$, and $\ve{n}_{m}$ at the positions of the pilot subcarriers, and let the diagonal matrix $ \mh{H}^\text{p}\in \mathbb{C}^{N_\text{p} \times N_\text{p}}$ be the sub-matrix of $\mh{H}$ containing the channel coefficients for the pilot subcarriers only\footnote{In the remainder of this work, the superscript 'p' indicates a sub-vector/matrix containing only the elements corresponding to pilot symbols/subcarriers.}. Based on the model in \eqref{equ:IEEE_RDM_data001}, they are connected via
\begin{align}
	\ve{z}_{m}^\text{p} &\approx{}   \mh{H}^\text{p}  \ve{s}_{m}^\text{p} \text{e}^{\mathrm{j} \varphi_m}  + \ve{n}^\text{p}_{m} \\
	&={}   \mh{H}^\text{p}  \ve{x}_{m}^\text{p}  + \ve{n}^\text{p}_{m}, \label{equ:IEEE_RDM_data001a} 
\end{align}
where we replaced the true channel with its estimate, and where $\ve{x}_{m}^\text{p} = \ve{s}_{m}^\text{p} \text{e}^{\mathrm{j} \varphi_m}$ represents the \ac{cpe} distorted pilot symbols. Applying the \ac{lmmse} estimator \cite{Kay-Est.Theory, huemer2011classical, Diss_Hofbauer, Diss_Lang_Oliver} for estimating $\ve{x}_{m}^\text{p}$ yields  
\begin{align}
\widehat{\ve{x}}^\text{p}_{m} &={} \left( \left(\mh{H}^\text{p}\right)^H \mh{H}^\text{p} + N_\text{c} \sigma_n^2 \left(\m{C}^\text{p}_{\ve{x}\ve{x}}\right)^{-1} \right)^{-1}  \left(\mh{H}^\text{p}\right)^H \ve{z}_{m}^\text{p},	 \label{equ:IEEE_RDM_CIR032e} 
\end{align}
whose error covariance matrix is given by
\begin{align}
\m{C}^\text{p}_{\ve{e}\ve{e}} &={} N_\text{c} \sigma_n^2 \left( \left(\mh{H}^\text{p}\right)^H \mh{H}^\text{p} + N_\text{c} \sigma_n^2  \left(\m{C}^\text{p}_{\ve{x}\ve{x}}\right)^{-1} \right)^{-1}.	 \label{equ:IEEE_RDM_pilot006} 
\end{align}
Here, $\m{C}^\text{p}_{\ve{x}\ve{x}} \in \mathbb{C}^{N_\text{p} \times N_\text{p}}$ denotes the covariance matrix of $\ve{x}^\text{p}_{m}$, which is assumed to be a diagonal matrix whose elements are given by the average power (averaged over $m$) of the pilot subcarrier symbols in $\ve{s}^\text{p}_{m}$.
Based on the estimate $\widehat{\ve{x}}^\text{p}_{m}$ in \eqref{equ:IEEE_RDM_CIR032e}, the \ac{cpe} is estimated via \cite{classen1994frequency, Diss_Hofbauer, huemer2002simulation}
\begin{align}
\widehat{\varphi}_m = \text{arg} \left( \left( \ve{s}_{m}^\text{p} \right)^H \left( \m{C}^\text{p}_{\ve{e}\ve{e}} \right)^{-1} \, \widehat{\ve{x}}_{m}^\text{p} \right).	 \label{equ:IEEE_RDM_a_pilot006a} 
\end{align}
Let $\ve{z}_{m}^\text{d}\in \mathbb{C}^{N_\text{d}}$, $\ve{s}_{m}^\text{d}\in \mathbb{C}^{N_\text{d}}$, and $\ve{n}_{m}^\text{d}\in \mathbb{C}^{N_\text{d}}$ be the sub-vectors of their respective signal vectors $\ve{z}_{m}$, $\ve{s}_{m}$, and $\ve{n}_{m}$ at the positions of the data subcarriers, and let the diagonal matrix $ \mh{H}^\text{d}\in \mathbb{C}^{N_\text{d} \times N_\text{d}}$ be the sub-matrix of $\mh{H}$ containing the channel coefficients for the data subcarriers only\footnote{In the remainder of this work, the superscript 'd' indicates a sub-vector/matrix containing only the elements corresponding to data symbols/subcarriers.}. With these definitions, a modification of the model in \eqref{equ:IEEE_RDM_data001} yields
\begin{align}
	\widetilde{\ve{z}}^\text{d}_{m} &={} \ve{z}^\text{d}_{m} \, \text{e}^{-\mathrm{j} \widehat{\varphi}_m} \in \mathbb{C}^{N_\text{d}} \\
	&\approx{}  \mh{H}^\text{d}  \ve{s}^\text{d}_{m} + \ve{n}^\text{d}_{m} \text{e}^{-\mathrm{j} \widehat{\varphi}_m} \label{equ:IEEE_RDM_data001y} 
\end{align}
for $m \geq N_\text{pr}$, which represents the synchronized receive vectors. 

\vspace{1mm} 

\noindent \textbf{Synchronization for FRIQIR OFDM}:
 The synchronization procedure for \ac{iqir} \ac{ofdm}  follows the same processing steps as for \ac{iqirSotA} \ac{ofdm} when
\begin{enumerate}
\item replacing the matrix $\mh{H}^\text{p} $ by $\mh{H}^\text{p}_1$ and $\mh{H}^\text{p}_2$ for even and odd values of $m$, respectively, and
\item replacing the matrix $\mh{H}^\text{d} $ by $\mh{H}^\text{d}_1$ and $\mh{H}^\text{d}_2$ for even and odd values of $m$, respectively.
\end{enumerate}
Then, the result of the synchronization procedure is similar to that in \eqref{equ:IEEE_RDM_data001y} and given for $m \geq N_\text{pr}$ by
\begin{align}
	\widetilde{\ve{z}}^\text{d}_{m} &={}  \mh{H}_1^\text{d}  \ve{s}^\text{d}_{m} + \ve{n}^\text{d}_{m} \text{e}^{-\mathrm{j} \widehat{\varphi}_m} \label{equ:IEEE_RDM_data001yeven} 
\end{align}
for even values of $m$, and
\begin{align}
	\widetilde{\ve{z}}^\text{d}_{m} &={}  \mh{H}_2^\text{d}  \ve{s}^\text{d}_{m} + \ve{n}^\text{d}_{m} \text{e}^{-\mathrm{j} \widehat{\varphi}_m} \label{equ:IEEE_RDM_data001yodd} 
\end{align}
for odd values of $m$.

\subsection{Data Estimation} \label{sec:Data_Estimation}


\vspace{1mm} 

\noindent \textbf{Data estimation for standard OFDM}:
For this communication system, the data symbols are estimated using the standard \ac{lmmse} estimator \cite{Kay-Est.Theory, huemer2011classical, huemer2002simulation, classen1994frequency, Diss_Hofbauer, Lang_RDM_JP, Van_Nee_OFDM, salehi2007digital, haykin2013digital, Lang_Asilomar_2016_LLRs, allpress2004exact} applied on the receive vector $\ve{z}_{m}$ in \eqref{equ:IEEE_RDM_data001}. We note that for standard \ac{ofdm} the \ac{lmmse} estimator attains the same \ac{ber} performance as the zero-forcing estimator. 

\vspace{1mm} 

\noindent \textbf{Data estimation for \ac{iqirSotA} \ac{ofdm}}: 
Basis of data estimation is the synchronized receive vector $\widetilde{\ve{z}}^\text{d}_{m}$ in \eqref{equ:IEEE_RDM_data001y} except that the elements corresponding to positive and negative indices $k$ are treated differently. Let  $ \widetilde{\ve{z}}^\text{d,+}_{m}\in \mathbb{C}^{N_\text{d}/2}$ be the sub-vector of $\widetilde{\ve{z}}^\text{d}_{m}$ containing the values for positive values of $k$, and let $ \widetilde{\ve{z}}^\text{d,-}_{m}\in \mathbb{C}^{N_\text{d}/2}$ be the sub-vector of $\widetilde{\ve{z}}^\text{d}_{m}$ containing the values for negative values of $k$ such that
\begin{align}
	\widetilde{\ve{z}}^\text{d}_{m} &={} \begin{bmatrix}
	\widetilde{\ve{z}}^\text{d,-}_{m} \\ \widetilde{\ve{z}}^\text{d,+}_{m}
\end{bmatrix}.	 \label{equ:IEEE_RDM_data001xy} 
\end{align}
Separating $\mh{H}^\text{d} $, $ \ve{s}^\text{d}_{m}$, and $ \ve{n}^\text{d}_{m} $ allows rewriting \eqref{equ:IEEE_RDM_data001y} as
\begin{align}
	\begin{bmatrix}
	\widetilde{\ve{z}}^\text{d,-}_{m} \\ \widetilde{\ve{z}}^\text{d,+}_{m}
\end{bmatrix} =& \begin{bmatrix}
	\mh{H}^\text{d,-} & \m{0}^{N_\text{d}/2 \times N_\text{d}/2} \\ \m{0}^{N_\text{d}/2 \times N_\text{d}/2} &\mh{H}^\text{d,+}
\end{bmatrix} \, \begin{bmatrix}
	\ve{s}^\text{d,-}_{m} \\ \ve{s}^\text{d,+}_{m}
\end{bmatrix}  + \begin{bmatrix}
	\ve{n}^\text{d,-}_{m} \\ \ve{n}^\text{d,+}_{m}
\end{bmatrix} \text{e}^{-\mathrm{j} \widehat{\varphi}_m}. \label{equ:IEEE_RDM_data020} 
\end{align}
The following reformulations bring the model in \eqref{equ:IEEE_RDM_data020} in a form that only depends on the data symbols with positive values of $k$, $\ve{s}^\text{d,+}_{m}$. Note that $\ve{s}^\text{d,-}_{m} $ and $ \ve{s}^\text{d,+}_{m}$ do not contain the real-valued subcarrier symbols for  $k \in \{ -N_\text{c}/2,\,\, 0  \}$ as they are used as pilots according to Sec.~\ref{sec:Signal_Model}. Thus, the connection between these two vectors as defined by the design rule I can be formulated as
\begin{align}
	\ve{s}^\text{d,-}_{m} &={} \m{D}^\text{I} \m{P}  \left(\ve{s}^\text{d,+}_{m} \right)^*, \label{equ:IQIR_OFDM_Comm031} 
\end{align}
where $\m{P} \in \mathbb{R}^{N_\text{d}/2 \times N_\text{d}/2}$ is a permutation matrix which reverses the order of the elements in $\ve{s}^\text{d,+}_{m}$, and where $\m{D}^\text{I} \in \mathbb{C}^{N_\text{d}/2 \times N_\text{d}/2}$ is a diagonal matrix that applies the phase rotations $\text{e}^{\text{j} \pi k}$. Combining \eqref{equ:IEEE_RDM_data020} and \eqref{equ:IQIR_OFDM_Comm031} yields
\begin{align}
	\begin{bmatrix}
	\widetilde{\ve{z}}^\text{d,-}_{m} \\ \widetilde{\ve{z}}^\text{d,+}_{m}
\end{bmatrix} \hspace{-1mm} &={} \hspace{-1mm} \begin{bmatrix}
	\mh{H}^\text{d,-} \m{D}^\text{I} \m{P} & \m{0}^{N_\text{d}/2 \times N_\text{d}/2} \\ \m{0}^{N_\text{d}/2 \times N_\text{d}/2} &\mh{H}^\text{d,+}
\end{bmatrix} \hspace{-1mm} \begin{bmatrix}
	\left(\ve{s}^\text{d,+}_{m} \right)^* \\ \ve{s}^\text{d,+}_{m}
\end{bmatrix}\hspace{-1mm} + \hspace{-1mm} \begin{bmatrix}
	\ve{n}^\text{d,-}_{m} \\ \ve{n}^\text{d,+}_{m}
\end{bmatrix} \text{e}^{-\mathrm{j} \widehat{\varphi}_m}. \label{equ:IEEE_RDM_data021} 
\end{align}
Taking the complex conjugate of $\widetilde{\ve{z}}^\text{d,-}_{m}$ yields
\begin{align}
	\begin{bmatrix}
	\left(\widetilde{\ve{z}}^\text{d,-}_{m}\right)^* \\ \widetilde{\ve{z}}^\text{d,+}_{m}
\end{bmatrix} =& \begin{bmatrix}
	\left(\mh{H}^\text{d,-} \m{D}^\text{I} \m{P} \right)^* & \m{0}^{N_\text{d}/2 \times N_\text{d}/2} \\ \m{0}^{N_\text{d}/2 \times N_\text{d}/2} &\mh{H}^\text{d,+}
\end{bmatrix} \, \begin{bmatrix}
	\ve{s}^\text{d,+}_{m}  \\ \ve{s}^\text{d,+}_{m}
\end{bmatrix}  \nonumber \\
&+ \begin{bmatrix}
	\left( \ve{n}^\text{d,-}_{m} \right)^* \text{e}^{\mathrm{j} \widehat{\varphi}_m} \\  \ve{n}^\text{d,+}_{m} \text{e}^{-\mathrm{j} \widehat{\varphi}_m} \end{bmatrix}  \\
 =& \begin{bmatrix}
	\left(\mh{H}^\text{d,-} \m{D}^\text{I} \m{P} \right)^*  \\ \mh{H}^\text{d,+}
\end{bmatrix} \, \ve{s}^\text{d,+}_{m} + \begin{bmatrix}
	\left( \ve{n}^\text{d,-}_{m} \right)^* \text{e}^{\mathrm{j} \widehat{\varphi}_m} \\  \ve{n}^\text{d,+}_{m} \text{e}^{-\mathrm{j} \widehat{\varphi}_m} \end{bmatrix} . \label{equ:IEEE_RDM_data022} 
\end{align}
With 
\begin{align}
&\widetilde{\ve{z}}^\text{d,I}_{m} = 	\begin{bmatrix}
	\left(\widetilde{\ve{z}}^\text{d,-}_{m}\right)^* \\ \widetilde{\ve{z}}^\text{d,+}_{m}
\end{bmatrix}, \hspace{5mm} \mh{H}^\text{d,I} =  \begin{bmatrix}
	\left(\mh{H}^\text{d,-} \m{D}^\text{I} \m{P} \right)^*  \\ \mh{H}^\text{d,+}
\end{bmatrix}, \nonumber \\
& \ve{n}^\text{d,I}_{m} =\begin{bmatrix}
	\left( \ve{n}^\text{d,-}_{m} \right)^* \text{e}^{\mathrm{j} \widehat{\varphi}_m} \\  \ve{n}^\text{d,+}_{m} \text{e}^{-\mathrm{j} \widehat{\varphi}_m} \end{bmatrix}, \label{equ:IEEE_RDM_data023} 
\end{align}
the model in \eqref{equ:IEEE_RDM_data022} simplifies to
\begin{align}
	\widetilde{\ve{z}}^\text{d,I}_{m} =&  \mh{H}^\text{d,I} \ve{s}^\text{d,+}_{m} + \ve{n}^\text{d,I}_{m}. \label{equ:IEEE_RDM_data024} 
\end{align}
Based on that, the data symbols in $\ve{s}^\text{d,+}_{m}$ can be estimated using the standard \ac{lmmse} estimator\footnote{Note that for standard OFDM, the CPEs are estimated based on the output of the LMMSE data estimator. This is not possible for IQIR OFDM because the complex conjugate operation in \eqref{equ:IEEE_RDM_data022} and the simplification in \eqref{equ:IEEE_RDM_data024} distort the \acp{cpe}.} 
\begin{align}
\widehat{\ve{s}}^\text{d,+}_{m} &={} \left( \left(\mh{H}^\text{d,I}\right)^H \mh{H}^\text{d,I} + N_\text{c} \sigma_n^2 \left(\m{C}^\text{d,+}_{\ve{x}\ve{x}}\right)^{-1} \right)^{-1} \left(\mh{H}^\text{d,I}\right)^H \widetilde{\ve{z}}^\text{d,I}_{m},	 \label{equ:IEEE_RDM_data026} 
\end{align}
whose error covariance matrix is given by
\begin{align}
\m{C}^\text{d}_{\ve{e}\ve{e}} &={} N_\text{c} \sigma_n^2 \left( \left(\mh{H}^\text{d,I}\right)^H \mh{H}^\text{d,I} + N_\text{c} \sigma_n^2  \left(\m{C}^\text{d,+}_{\ve{x}\ve{x}}\right)^{-1} \right)^{-1},	 \label{equ:IEEE_RDM_data027} 
\end{align}
where $\m{C}^\text{d,+}_{\ve{x}\ve{x}}\in \mathbb{C}^{N_\text{d}/2 \times N_\text{d}/2}$ is the covariance matrix of $\ve{s}^\text{d,+}_{m}$.

\vspace{1mm} 

\noindent \textbf{Data estimation for FRIQIR OFDM}:
For this communication system, data estimation follows in principle the same steps as for \ac{iqirSotA} \ac{ofdm} except for:
\begin{enumerate}
\item The matrix $\mh{H}^\text{d} $ needs to be replaced by $\mh{H}^\text{d}_1$ and $\mh{H}^\text{d}_2$ for even and odd values of $m$, respectively. 
\item $\m{D}^\text{I} $ in \eqref{equ:IQIR_OFDM_Comm031} needs to be replaced by the $\m{D}^{\text{II},m} = \text{e}^{-\text{j} \pi m} \m{I}^{N_\text{d}/2} \in \mathbb{C}^{N_\text{d}/2 \times N_\text{d}/2}$ to reverse the artificial Doppler shift $\text{e}^{\text{j} \pi m}$ in design rule II. 
\end{enumerate}

\section{BER Simulation Setup and Performance Comparison} \label{sec:BER_performance}

The \ac{ber} performances of all three communication systems described previously in the presence of \ac{iq} imbalance are compared in this section. 
Please note that for standard \ac{ofdm}, numerous algorithms exist to cancel \ac{iq} imbalance in the communication receiver. However, a comparison with these algorithms is beyond the scope of this work and will be topic of future research. Instead, we just observe the degrading effects of \ac{iq} imbalance on the considered systems.
The parametrization and the \ac{iq} imbalance model correspond to that used in Sec.~\ref{sec:Testing_SIM}. Additional processing blocks not detailed in Sec.~\ref{sec:Comm_System} correspond to that employed in \cite{Lang_SA_OFDM, Lang_RDM_JP} and are repeated in the following for the sake of completeness.

At the transmitter, we employ a channel encoder, an interleaver, and a mapper for all communication systems. The channel encoder with coding rate of $r = 1/2$ applies a convolutional channel code with generator polynomial $(133, 171)_8$ in octal representation  and  constraint length 7 \cite{Diss_Hofbauer, IEEE99}  on the bitstream. The coded bitstream is fed into a random interleaver with a block length equal to the number of coded bits in an \ac{ofdm} symbol, followed by a mapper that transforms the bits into \ac{qpsk} subcarrier symbols.

The model employed for generating random channels $\ve{f}$ is described in \cite{ Rappaport_SISO_Channel_JP, Rappaport_SISO_Channel} and code for automated generation of \acp{cir} is available in \cite{Code_MIMO_Channels}. With this code, in which we used the very same parametrization as detailed in \cite{Lang_RDM_JP},   $5\,000$ \ac{los} channels and $5\,000$ \ac{nlos} channels were generated and used for averaging the \ac{ber} performance. For the channel model used in Sec.~\ref{sec:BER_performance}, a statistical evaluation based on a large set of generated \acp{cir} confirmed $N_\ve{f} = 256$ to be sufficient for adequate modeling \cite{Lang_RDM_JP}. For every channel realization, a relative velocity between transmitter and receiver is chosen randomly from a uniform distribution between $\pm 60\, \text{m/s}$ and is kept constant for the duration of a burst with $N_\text{sym}$ \ac{ofdm} symbols. This relative velocity causes distortions in form of \ac{ici} and a \ac{cpe}\footnote{In this work, we estimate the effective channels based on $N_\text{pr}$ preamble \ac{ofdm} symbols transmitted at the beginning of every burst. However, we note that for the assumed relative velocity between $\pm 60\,\text{m/s}$ the channels may become highly time-varying in practice, entailing the necessity of much more frequent channel estimation or advanced channel tracking algorithms. A detailed investigation of these effects or an analysis of such algorithms is beyond the scope of this work.}. Following state-of-the-art approaches, the synchronization methods were limited to the \ac{cpe} compensation while \ac{ici} remained uncompensated.

At the receiver side, the estimated subcarrier symbols from the \ac{lmmse} output are fed into a demapper, which derives the soft information in form of \acp{llr} \cite{allpress2004exact, Haselmayr_LLRs, Lang_Asilomar_2016_LLRs}, which are then deinterleaved and decoded using a Viterbi decoder \cite{Viterbi}.

\Ac{awgn} with $\mathcal{N}\left(0,\, \sigma_\text{n}^2\right)$  is added to the signal at the receiver input. The variance $\sigma_\text{n}^2$ is set to attain a chosen $E_\text{b}/N_0$ value with $E_\text{b}$ denoting the average energy per bit of information, and with $N_0/2$ representing the double-sided noise power spectral density of the bandpass noise \cite{Diss_Hofbauer}. This noise variance is set according to \cite{Miller_compl_BB}
\begin{align}
	\sigma_\text{n}^2 = \frac{P_\text{s}}{(E_\text{b}/N_0) \,r\, b\, \zeta \, \nu }. \label{equ:IEEE_CIR015a}	
\end{align}
Therein, $P_\text{s}$ is the average time domain power at the receiver input, and $b$ denotes the number of coded bits per data symbol, which is set to $2$, since \ac{qpsk} symbols were used in the simulations.  $\zeta$ in \eqref{equ:IEEE_CIR015a} accounts for the samples in the \ac{cp} and is set to $\zeta = N_\text{c}/(N_\text{cp}+N_\text{c})$. 
$\nu$ is a parameter that accounts for the fact that \ac{iqirSotA} \ac{ofdm}, and \ac{iqir} \ac{ofdm} introduce redundancy to the transmitted subcarrier symbols in $\m{S}$.
Hence, $\nu$ is set to $1/2$ for these two communication systems, and to $1$ for the standard \ac{ofdm}. Consequently, the standard \ac{ofdm}'s noise variance in the simulations is lower than for the other communication systems.

\subsubsection{Perfect Channel Knowledge; Perfect CPE Synchronization}

\begin{figure}[!t]
\begin{center}
\begin{tikzpicture}
\begin{axis}[compat=newest, 
width=.95\columnwidth, height = .5\columnwidth, grid, xlabel={$E_b/N_0$ (dB)}, 
ylabel={$\text{log}_{10}(\text{BER})$}, 
legend pos=south east, 
legend cell align=left,
legend columns=3, 
xmin = 0,
xmax = 16,
ymin = -6,
ymax = 0,
ytick={0, -1, -2, -3, -4, -5, -6},
legend style={
at={(-0.1, 1.28)},
anchor=north west,font=\tiny}
]

\addplot[line width=1pt][color=black, solid] table[x index =0, y index =1] {SimData/IQ_SISO_Nc512_Nsym256_R12_CK0_CD2_Nbps2_Nprea8_Npi8_IQTx1Rx5_ZINcm100.dat};
\addlegendentry{{\footnotesize Std. OFDM}}

\addplot[line width=1pt][color=gray, solid, every mark/.append style={solid},mark=star,mark repeat = 2] table[x index =0, y index =1] {SimData/IQ_SotA_Nc512_Nsym256_R12_CK0_CD2_Nbps2_Nprea16_Npi16_IQTx1Rx5_ZINcm100.dat};
\addlegendentry{{\footnotesize IQIR OFDM}}

\addplot[line width=1pt][color=lightgray, solid] table[x index =0, y index =1] {SimData/IQ_Prop_Nc512_Nsym256_R12_CK0_CD2_Nbps2_Nprea16_Npi16_IQTx1Rx5_ZINcm100.dat};
\addlegendentry{{\footnotesize FRIQIR OFDM}}

\addplot[line width=1pt][color=black, dashed] table[x index =0, y index =1] {SimData/IQ_SISO_Nc512_Nsym256_R12_CK0_CD2_Nbps2_Nprea8_Npi8_IQTx0Rx0_ZINcm100.dat};

\addplot[line width=1pt][color=gray, dashed, every mark/.append style={solid},mark=star,mark repeat = 2] table[x index =0, y index =1] {SimData/IQ_SotA_Nc512_Nsym256_R12_CK0_CD2_Nbps2_Nprea8_Npi8_IQTx0Rx0_ZINcm100.dat};

\addplot[line width=1pt][color=lightgray, dashed] table[x index =0, y index =1] {SimData/IQ_Prop_Nc512_Nsym256_R12_CK0_CD2_Nbps2_Nprea8_Npi8_IQTx0Rx0_ZINcm100.dat};

\end{axis}
\end{tikzpicture}
\caption{BER performance for perfect IQ balanced (dashed) and for IQ imbalance distorted (solid) communications. The effective channels are assumed to be known and the CPE is compensated perfectly. The dashed lines for IQIR OFDM and  FRIQIR OFDM lie on top of each other.
\label{fig:BER_optimal} }
\end{center}
\end{figure}
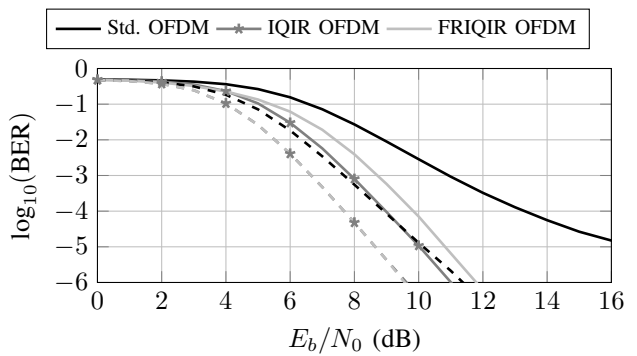

We first investigate the \ac{ber} performance degradation due to \ac{iq} imbalance. The effective channels are assumed to be known, and the \ac{cpe} is compensated perfectly. Fig.~\ref{fig:BER_optimal} shows the resulting \ac{ber} curves for the cases of perfect \ac{iq} balance (dashed), and with \ac{iq} imbalance (solid). A comparison reveals that the loss in \ac{ber} performance due to \ac{iq} imbalance is severe for standard \ac{ofdm}. In contrast to that, this loss is only $ \sim 1.5 \, \text{dB}$ and only $ \sim 2 \, \text{dB}$ for \ac{iqirSotA} \ac{ofdm} and \ac{iqir} \ac{ofdm}, respectively. 


\subsubsection{Perfect CPE Synchronization; Imperfect Channel Knowledge}

Next, the \ac{ber} performance degradation when estimating the effective channels is investigated. For that, the effective channels for \ac{iqirSotA} \ac{ofdm} as well as for \ac{iqir} \ac{ofdm} are estimated based on $N_\text{pr} = 16$ preamble \ac{ofdm} symbols. For standard \ac{ofdm}, $N_\text{pr}$ is reduced to $8$. This reduction is conducted to ensure the same effective \ac{snr} when averaging over the $N_\text{pr}$ preamble \ac{ofdm} symbols \cite{Lang_RDM_JP, Lang_SA_OFDM}. 

\begin{figure}[!t]
\begin{center}
\begin{tikzpicture}
\begin{axis}[compat=newest, 
width=.95\columnwidth, height = .5\columnwidth, grid, xlabel={$E_b/N_0$ (dB)}, 
ylabel={$\text{log}_{10}(\text{BER})$}, 
legend pos=south east, 
legend cell align=left,
legend columns=3, 
xmin = 0,
xmax = 16,
ymin = -6,
ymax = 0,
ytick={0, -1, -2, -3, -4, -5, -6},
legend style={
at={(-0.1, 1.28)},
anchor=north west,font=\tiny}
]

\addplot[line width=1pt][color=black, solid] table[x index =0, y index =1] {SimData/IQ_SISO_Nc512_Nsym256_R12_CK0_CD2_Nbps2_Nprea8_Npi8_IQTx1Rx5_ZINcm100.dat};
\addlegendentry{{\footnotesize Std. OFDM}}

\addplot[line width=1pt][color=gray, solid, every mark/.append style={solid},mark=star,mark repeat = 2] table[x index =0, y index =1] {SimData/IQ_SotA_Nc512_Nsym256_R12_CK0_CD2_Nbps2_Nprea16_Npi16_IQTx1Rx5_ZINcm100.dat};
\addlegendentry{{\footnotesize IQIR OFDM}}

\addplot[line width=1pt][color=lightgray, solid] table[x index =0, y index =1] {SimData/IQ_Prop_Nc512_Nsym256_R12_CK0_CD2_Nbps2_Nprea16_Npi16_IQTx1Rx5_ZINcm100.dat};
\addlegendentry{{\footnotesize FRIQIR OFDM}}

\addplot[line width=1pt][color=black, dashed] table[x index =0, y index =1] {SimData/IQ_SISO_Nc512_Nsym256_R12_CK1_CD2_Nbps2_Nprea8_Npi8_IQTx1Rx5_ZINcm100.dat};

\addplot[line width=1pt][color=gray, dashed, every mark/.append style={solid},mark=star,mark repeat = 2] table[x index =0, y index =1] {SimData/IQ_SotA_Nc512_Nsym256_R12_CK1_CD2_Nbps2_Nprea16_Npi16_IQTx1Rx5_ZINcm100.dat};

\addplot[line width=1pt][color=lightgray, dashed] table[x index =0, y index =1] {SimData/IQ_Prop_Nc512_Nsym256_R12_CK1_CD2_Nbps2_Nprea16_Npi16_IQTx1Rx5_ZINcm100.dat};

\end{axis}
\end{tikzpicture}
\caption{BER performance for the case of perfect CPE synchronization. The solid lines arise as a result of perfect channel knowledge, and
the dashed lines from imperfectly estimated channels based on preamble OFDM symbols.
\label{fig:BER_channel_est} }
\end{center}
\end{figure}
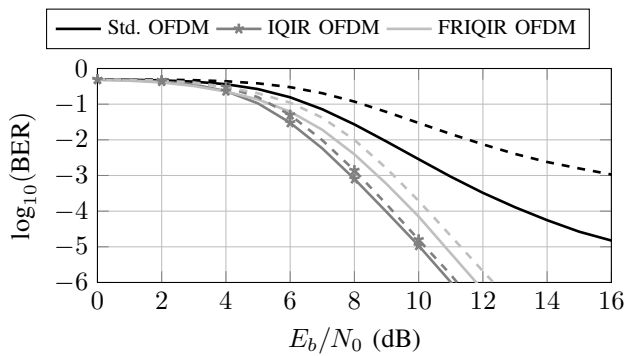

The resulting \ac{ber} curves are shown in Fig.~\ref{fig:BER_channel_est} (dashed lines) together with the results obtained for perfect channel knowledge from the previous simulation (solid lines). This figure reveals that the loss in \ac{ber} performance due to imperfect channel knowledge is small for  \ac{iqirSotA} \ac{ofdm} and  \ac{iqir} \ac{ofdm}. In contrast to that, the \ac{ber} performance of standard \ac{ofdm} is severely degraded.

\subsubsection{Perfect Channel Knowledge; Imperfect CPE Synchronization}

For this simulation, the effective channels are assumed to be perfectly known and \ac{cpe} synchronization is performed based on pilot subcarriers known to the receivers. The number of pilot subcarriers is set to $N_\text{p} = 16$ for \ac{iqirSotA} \ac{ofdm} and \ac{iqir} \ac{ofdm}. Due to the same argumentation as for the previous simulations, we reduce the number of pilot subcarriers to $N_\text{p} = 8$ for standard \ac{ofdm} such that the same effective \ac{snr} is obtained for the estimated \ac{cpe}. 

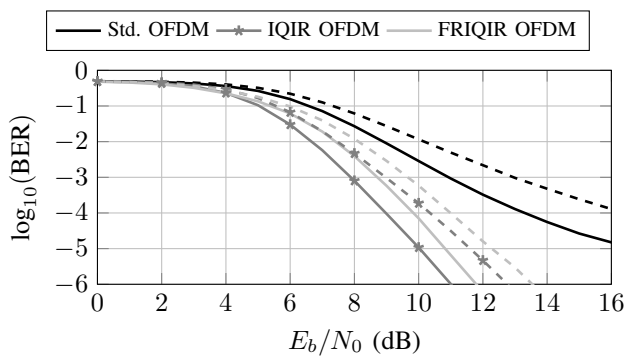
\begin{figure}[!t]
\begin{center}
\begin{tikzpicture}
\begin{axis}[compat=newest, 
width=.95\columnwidth, height = .5\columnwidth, grid, xlabel={$E_b/N_0$ (dB)}, 
ylabel={$\text{log}_{10}(\text{BER})$}, 
legend pos=south east, 
legend cell align=left,
legend columns=3, 
xmin = 0,
xmax = 16,
ymin = -6,
ymax = 0,
ytick={0, -1, -2, -3, -4, -5, -6},
legend style={
at={(-0.1, 1.28)},
anchor=north west,font=\tiny}
]

\addplot[line width=1pt][color=black, solid] table[x index =0, y index =1] {SimData/IQ_SISO_Nc512_Nsym256_R12_CK0_CD2_Nbps2_Nprea8_Npi8_IQTx1Rx5_ZINcm100.dat};
\addlegendentry{{\footnotesize Std. OFDM}}

\addplot[line width=1pt][color=gray, solid, every mark/.append style={solid},mark=star,mark repeat = 2] table[x index =0, y index =1] {SimData/IQ_SotA_Nc512_Nsym256_R12_CK0_CD2_Nbps2_Nprea16_Npi16_IQTx1Rx5_ZINcm100.dat};
\addlegendentry{{\footnotesize IQIR OFDM}}

\addplot[line width=1pt][color=lightgray, solid] table[x index =0, y index =1] {SimData/IQ_Prop_Nc512_Nsym256_R12_CK0_CD2_Nbps2_Nprea16_Npi16_IQTx1Rx5_ZINcm100.dat};
\addlegendentry{{\footnotesize FRIQIR OFDM}}

\addplot[line width=1pt][color=black, dashed] table[x index =0, y index =1] {SimData/IQ_SISO_Nc512_Nsym256_R12_CK0_CD1_Nbps2_Nprea8_Npi8_IQTx1Rx5_ZINcm100.dat};

\addplot[line width=1pt][color=gray, dashed, every mark/.append style={solid},mark=star,mark repeat = 2] table[x index =0, y index =1] {SimData/IQ_SotA_Nc512_Nsym256_R12_CK0_CD1_Nbps2_Nprea16_Npi16_IQTx1Rx5_ZINcm100.dat};

\addplot[line width=1pt][color=lightgray, dashed] table[x index =0, y index =1] {SimData/IQ_Prop_Nc512_Nsym256_R12_CK0_CD1_Nbps2_Nprea16_Npi16_IQTx1Rx5_ZINcm100.dat};

\end{axis}
\end{tikzpicture}
\caption{BER performance for the case of perfect channel knowledge. The solid lines arise as a result of perfect CPE synchronization,
and the dashed lines from imperfect CPE synchronization based on pilot subcarriers.
\label{fig:BER_sync} }
\end{center}
\end{figure}

The obtained \ac{ber} curves are shown in Fig.~\ref{fig:BER_sync} (dashed lines). This figure also contains the \ac{ber} curves for the case of perfect \ac{cpe} synchronization from the first simulation. A comparison indicates that the loss in \ac{ber} performance due to imperfect \ac{cpe} synchronization for  \ac{iqirSotA} \ac{ofdm} is approximately the same as for  \ac{iqir} \ac{ofdm}.

We summarize the simulation results by stating that with \ac{iqirSotA} \ac{ofdm} and with \ac{iqir} \ac{ofdm}, \ac{iq} imbalance may can be neglected due to its minor influence on the \ac{ber} performance, while for standard \ac{ofdm}, \ac{iq} imbalance cancellation methods might be necessary to achieve proper performance.

\section{Conclusion} \label{sec:Conclusions}
In this work, a novel \ac{ofdm}-based waveform for joint sensing and communication denoted as \ac{iqir} \ac{ofdm} was presented. This waveform was designed to be robust against \ac{iq} imbalance by means of avoiding an increased noise floor, which was achieved by adding redundancy to the transmit data according to a defined design rule. As for the \ac{iqirSotA} \ac{ofdm} waveform proposed in \cite{IQ_Imbalance_Rad2}, \ac{iq} imbalance causes up to three ghost objects in the \ac{rvm} for every real object. For \ac{iqir} \ac{ofdm}, different to \cite{IQ_Imbalance_Rad2}, these ghost objects are located in the \ac{rvm} at the same range bin as the real object, but at different velocity bins. Hence, their changes in range observed through a comparison of several consecutive \acp{rvm} are not as expected for a real object, since it does not correspond to the observed velocity. This enables tracking algorithms to identify them as ghost objects, and avoids a reduction of the maximum unambiguous range as was necessary for \ac{iqirSotA} \ac{ofdm}.



We additionally analyzed the proposed \ac{iqir} \ac{ofdm} waveform in the context of wireless communications. There, the so-called effective channels, i.e., the combination of the true channel and the distortions caused by \ac{iq} imbalance, were analyzed. It turned out that the effective channels for \ac{iqir} \ac{ofdm} differ for even and odd \ac{ofdm} symbol indices even for static propagation channels. This work also contains a similar analysis of the effective channel for the \ac{iqirSotA} \ac{ofdm} waveform, since the communication aspects of this waveform were not analyzed in \cite{IQ_Imbalance_Rad2}. Based on this analysis, we proposed suitable methods for channel estimation, synchronization, and data estimation for \ac{iqir} \ac{ofdm}- and \ac{iqirSotA} \ac{ofdm}-based communication systems that deal with the effective channels and that utilize the redundancy in the transmit data efficiently. The effectiveness of these communication systems were demonstrated in form of \ac{ber} simulations, revealing a significant gain in performance compared to standard \ac{ofdm} in the presence of \ac{iq} imbalance.


%


%
%
%





\bibliographystyle{IEEEtran}
\bibliography{References_DDM}

%








\end{document}